\documentclass[11pt]{article}
\DeclareMathAlphabet{\scr}{U}{rsfs}{m}{n}
\usepackage{latexsym}
\usepackage{epsfig}
\usepackage[mathscr]{eucal}
\usepackage{amsfonts}
\usepackage{amscd}
\usepackage{amsmath}
\usepackage{array}
\usepackage{amssymb}
\usepackage{colordvi}
\usepackage{enumerate}
\usepackage{graphicx}
\usepackage{booktabs}
\usepackage[footnotesize]{caption}
\usepackage{fancyhdr} 
\usepackage{pdfpages}
\usepackage{slashed}
\usepackage{tabularx}
\usepackage{longtable}
\usepackage{array}
\usepackage{hyperref}
\usepackage{relsize}
\usepackage{color}
\usepackage[merge,numbers,compress]{natbib}
\usepackage{rotating}

\hypersetup{
	colorlinks=true,        
	linkcolor=blue,          
	citecolor=black,        
	filecolor=red,      
	urlcolor=cyan           
}

\definecolor{ourbrown}{RGB}{155,100,15}
\definecolor{ourcyan}{RGB}{20,165,165}
\definecolor{ourpurple}{RGB}{145,0,140}
\definecolor{darkorange}{RGB}{225,100,0}
\definecolor{darkgreen}{RGB}{0,170,0}
\definecolor{darkgray}{RGB}{80,80,80}

\newcommand{\newc}{\newcommand}
\newc{\EW}{electroweak\;}
\newc{\DM}{dark matter\;}
\newc{\SM}{standard model\;}
\newc{\GEV}{\text{GeV}}
\newc{\KK}{Kaluza-Klein\;}
\newc{\ff}{fragmentation function\;}
\newc{\be}{\begin{equation}}
\newc{\ee}{\end{equation}}
\newc{\bi}{\begin{itemize}}
\newc{\ei}{\end{itemize}}
\newc{\benu}{\begin{enumerate}}
\newc{\eenu}{\end{enumerate}}
\newc{\bc}{\begin{center}}
\newc{\ec}{\end{center}}
\newc{\bfig}{\begin{figure}}
\newc{\efig}{\end{figure}}
\newc{\neutone}{\tilde{\chi}^0_1}
\newc{\sigmav}{\langle\sigma v \rangle}
\newc{\lamhs}{\lambda_{H\!S}}
\newc{\logJ}{\log (J_{40^\circ}/J_{40^\circ\!,\,\text{nom}})}
\newc{\sigJ}{\sigma_{\log\!J}}
\DeclareMathSymbol{\Gamma}{\mathalpha}{letters}{"00}
\DeclareMathSymbol{\Delta}{\mathalpha}{letters}{"01}
\DeclareMathSymbol{\Theta}{\mathalpha}{letters}{"02}
\DeclareMathSymbol{\Lambda}{\mathalpha}{letters}{"03}
\DeclareMathSymbol{\Xi}{\mathalpha}{letters}{"04}
\DeclareMathSymbol{\Pi}{\mathalpha}{letters}{"05}
\DeclareMathSymbol{\Sigma}{\mathalpha}{letters}{"06}
\DeclareMathSymbol{\Upsilon}{\mathalpha}{letters}{"07}
\DeclareMathSymbol{\Phi}{\mathalpha}{letters}{"08}
\DeclareMathSymbol{\Psi}{\mathalpha}{letters}{"09}
\DeclareMathSymbol{\Omega}{\mathalpha}{letters}{"0A}
\DeclareMathSymbol{\varGamma}{\mathalpha}{operators}{"00}
\DeclareMathSymbol{\varDelta}{\mathalpha}{operators}{"01}
\DeclareMathSymbol{\varTheta}{\mathalpha}{operators}{"02}
\DeclareMathSymbol{\varLambda}{\mathalpha}{operators}{"03}
\DeclareMathSymbol{\varXi}{\mathalpha}{operators}{"04}
\DeclareMathSymbol{\varPi}{\mathalpha}{operators}{"05}
\DeclareMathSymbol{\varSigma}{\mathalpha}{operators}{"06}
\DeclareMathSymbol{\varUpsilon}{\mathalpha}{operators}{"07}
\DeclareMathSymbol{\varPhi}{\mathalpha}{operators}{"08}
\DeclareMathSymbol{\varPsi}{\mathalpha}{operators}{"09}
\DeclareMathSymbol{\varOmega}{\mathalpha}{operators}{"0A}

\begin{document}

\title{\hfill ~\\[-30mm]
\phantom{h} \hfill\mbox{\small TTK-16-09} 
\\[1cm]
\vspace{13mm}   \textbf{
A global fit of the $\gamma$-ray galactic center excess within the scalar singlet Higgs portal model}
}
\date{March 27, 2016}
\author{
Alessandro Cuoco\footnote{E-mail: \texttt{cuoco@physik.rwth-aachen.de}}\;,
Benedikt Eiteneuer\footnote{E-mail: \texttt{eiteneuer@physik.rwth-aachen.de}}\;,
Jan Heisig\footnote{E-mail: \texttt{heisig@physik.rwth-aachen.de}}\;, and 
Michael Kr\"amer\footnote{E-mail: \texttt{mkraemer@physik.rwth-aachen.de}}\\[9mm]
{\it 
Institute for Theoretical Particle Physics and Cosmology,}\\ {\it RWTH Aachen University, 52056 Aachen, Germany}
}

\maketitle

\begin{abstract}
We analyse the excess in the $\gamma$-ray emission from the center of our galaxy 
observed by {\it Fermi}-LAT in terms of dark matter annihilation within the scalar Higgs 
portal model. In particular, we include the astrophysical uncertainties from the dark matter 
distribution and allow for unspecified additional dark matter components. We demonstrate 
through a detailed numerical fit that the strength and shape of the $\gamma$-ray spectrum 
can indeed be described by the model in various regions of dark matter masses and couplings. 
Constraints from invisible Higgs decays, direct dark matter searches, indirect searches in 
dwarf galaxies and for $\gamma$-ray lines, and constraints from the dark matter relic density 
reduce the parameter space to dark matter masses near the Higgs resonance. 
We find two viable regions: one where the Higgs-dark matter coupling is of ${\cal O}(10^{-2})$,  
and an additional dark matter component beyond the scalar WIMP of our model is preferred, 
and one region where the Higgs-dark matter coupling may be significantly smaller, but where the 
scalar WIMP constitutes a significant fraction or even all of dark matter. Both viable regions are 
hard to probe in future direct detection and collider experiments. 
\end{abstract}
\thispagestyle{empty}
\vfill
\newpage
\setcounter{page}{1}

\tableofcontents

\section{Introduction}

Weakly interacting massive particles (WIMPs) are promising candidates for dark matter (DM), and can be searched for 
at colliders and through direct and indirect detection experiments~\cite{Bertone:2004pz,Drees:2012ji,Klasen:2015uma}. The scalar singlet Higgs portal model is among the simplest  WIMP DM models. It comprises 
the Standard Model (SM) and a real singlet scalar DM field, $S$, which interacts 
with the SM Higgs field $H$ through the operator $S^2 H^\dagger H$~\cite{Silveira:1985rk,McDonald:1993ex,Burgess:2000yq}. The scalar Higgs portal model can 
accommodate the DM relic density, would contribute to the invisible Higgs width, and it can be 
probed in direct and indirect DM searches. 

An excess in the $\gamma$-ray emission 
from the center of our galaxy, as observed by the Large Area Telescope (LAT) on-board the {\it Fermi} satellite, has been reported by several groups in the last few years~\cite{Goodenough:2009gk,Hooper:2010mq,Hooper:2011ti,Abazajian:2012pn,Hooper:2013rwa,Gordon:2013vta,Abazajian:2014fta,Daylan:2014rsa,Calore:2014xka}, and has recently been confirmed by the {\it Fermi}-LAT collaboration~\cite{TheFermi-LAT:2015kwa}.
While there are various potential astrophysical explanations of such an excess, see e.g.~\cite{Petrovic:2014uda,Petrovic:2014xra,Mirabal:2012em,Yuan:2014rca,Bartels:2015aea,Lee:2015fea,Cholis:2015dea}, it is 
intriguing that the {\it Fermi}-LAT $\gamma$-ray spectrum and spatial distribution are consistent with a signal expected from DM annihilation~\cite{Goodenough:2009gk,Hooper:2010mq,Hooper:2011ti,Abazajian:2012pn,Hooper:2013rwa,Gordon:2013vta,Abazajian:2014fta,Daylan:2014rsa,Calore:2014xka,Alves:2014yha,Calore:2014nla,Agrawal:2014oha}. 
We will thus explore if the galactic center excess (GCE) can be explained in terms of DM annihilation within 
the minimal singlet scalar Higgs portal model, 
taking into account the constraints from invisible Higgs decays, direct DM searches, searches for DM annihilation from dwarf spheroidal galaxies, and 
searches for mono-energetic spectral $\gamma$-lines from the Milky Way halo.

As compared to previous Higgs portal model interpretations of the GCE~\cite{Agrawal:2014oha, Okada:2013bna,Basak:2014sza,Cline:2013gha,Duerr:2015bea}, we provide a detailed numerical fit of the GCE signal within the  scalar Higgs portal model, 
taking properly into account the theoretical uncertainty from the DM distribution. Furthermore, we allow for unspecified additional DM components 
beyond the scalar WIMP of our minimal model. We will show that, taking into account all the constraints, the scalar Higgs portal model can indeed 
describe the GCE signal, albeit only in a small region of parameter space near the Higgs resonance where the WIMP mass $m_S \sim m_h/2$. 

The paper is organised as follows. In section~\ref{sec:model} we introduce the scalar singlet Higgs portal model and briefly review previous 
collider and astroparticle analyses of this model. The DM annihilation $\gamma$-ray signatures of the scalar Higgs portal model 
are presented in section~\ref{sec:gce}, together with a discussion of the galactic center excess signal and the astrophysical uncertainties 
due to the dark matter distribution. We present a detailed numerical fit of the strength and shape of the GCE $\gamma$-ray spectrum, including 
in particular the astrophysical uncertainties and allowing for unspecified additional DM components. Constraints on the model parameters from the Higgs invisible width, direct 
detection searches, independent searches for $\gamma$-rays and from the dark matter relic density are discussed in section~\ref{sec:constraints}. In section~\ref{sec:results}  we finally present a global fit of the 
GCE within the scalar Higgs portal model, taking into account the above-mentioned constraints. We conclude in section~\ref{sec:summary}.

\section{The scalar singlet Higgs portal model}\label{sec:model}

The scalar singlet Higgs portal model~\cite{Silveira:1985rk,McDonald:1993ex,Burgess:2000yq} is among the simplest UV-complete WIMP DM models. The model comprises the Standard Model and a real scalar field, $S$, which is a singlet under all SM gauge groups. Imposing an additional $Z_2$ symmetry, $S \to -S$, the scalar particle is stable and thus a WIMP DM candidate. The Lagrangian of the scalar Higgs portal model reads 
\begin{equation}
{\cal L} = {\cal L}_\text{SM} + \frac 12 \partial_\mu  S  \partial^\mu  S  - \frac12 m_{S,0}^2S^2- \frac 14 \lambda_S  S^4- \frac 12 \lambda_{H\!S}\, S^2 H^\dagger H\,.
\label{eq:lagr}
\end{equation}
After electroweak symmetry breaking, the last three terms of the above Lagrangian become
\begin{equation}
{\cal L} \supset  - \frac12 m_{S}^2\, S^2- \frac 14 \lambda_S\,  S^4 - \frac 14 \lambda_{H\!S}\, h^2 S^2 - \frac {1}{2} \lambda_{H\!S}\, v h S^2\,,
\label{eq:ewbr}
\end{equation}
with $H = (h+v, 0)/\sqrt{2}\,$, $v = 246\,$GeV, and where we introduced the physical mass of the singlet 
field, $m_S^2  = m_{S,0}^2 + \lambda_{H\!S} v^2 / 2$. The scalar self coupling, $\lambda_S$, is of importance for the stability of the electroweak vacuum and the perturbativity of the model, see e.g.~\cite{Gonderinger:2009jp}, 
but does not affect DM phenomenology.\footnote{For an exception see e.g. \cite{Bernal:2015xba} 
where dark matter is strongly interacting.}
For the purpose of this paper, the model is thus fully specified by only two parameters beyond those of the SM:
the mass of the scalar DM particle, $m_S$, and the strength of the coupling between the DM and Higgs particles, 
$\lambda_{H\!S}$. 

The scalar singlet Higgs portal model defined in eq.~(\ref{eq:lagr}) is certainly minimal, and possibly too simplistic. However, a coupling between a new 
gauge singlet sector and the SM through the Higgs bilinear $H^\dagger H$ should be expected in a large class of SM extensions, as $H^\dagger H$ is the only SM gauge singlet operator of mass dimension two. Even within the minimal scalar Higgs portal model, eq.~(\ref{eq:lagr}), the $S^2 H^\dagger H$ interaction term gives rise to a rich phenomenology, including 
invisible Higgs decays, $h \to SS$, a DM-nucleon interaction through the exchange of a Higgs particle, and DM annihilation through 
$s$-channel Higgs, $t$-channel scalar exchange, and the $S^2 h^2$ interactions, see section~\ref{sec:gce}. 

The phenomenology of the singlet Higgs portal model has been extensively studied in the literature, see e.g.\ the recent reviews~\cite{Cline:2013gha,Beniwal:2015sdl} and references therein. Other recent general analyses 
of the model have been presented in \cite{Djouadi:2011aa,Cheung:2012xb}, while \cite{Djouadi:2012zc,Endo:2014cca,Craig:2014lda,Han:2016gyy} have specifically explored the constraints from searches at the Large Hadron Collider (LHC). Astrophysical constraints, in particular from $\gamma$-lines, have been studied in  \cite{Mambrini:2012ue, Feng:2014vea, Duerr:2015mva, Duerr:2015aka,Duerr:2015bea}. Constraints on the scalar Higgs portal model from perturbativity and electroweak vacuum stability have been revisited in \cite{Han:2015hda}, while the possibility to drive inflation through a non-minimal coupling of the scalar to gravity has been analysed in \cite{Kahlhoefer:2015jma} in light of current constraints. Extensions of the Higgs portal model that provide a similar 
phenomenology have been studied in \cite{Alvares:2012qv,Ghorbani:2014gka,Wang:2014elb,Kim:2016csm}.

\section{The galactic center excess}\label{sec:gce}

\subsection{The {\it Fermi}-LAT observation}

The presence of a GCE has been reported by several groups in 
the last few years~\cite{Goodenough:2009gk,Hooper:2010mq,Hooper:2011ti,Abazajian:2012pn,Hooper:2013rwa,Gordon:2013vta,Abazajian:2014fta,Daylan:2014rsa,Calore:2014xka}. 
The GCE seems compatible with a spherical morphology,  extending up to at least~10$^\circ$ away from the galactic center,   
and with a steep `cuspy' radial profile~\cite{Daylan:2014rsa,Calore:2014xka}.
The inferred energy spectrum  is peaked at a few GeV in the usual $E^2 \times$ flux representation.
Various astrophysical mechanisms and scenarios have been proposed to explain the excess~\cite{Petrovic:2014uda,Petrovic:2014xra,Cholis:2015dea}. 
On the other hand, intriguingly, it has been shown that the 
excess is also compatible with an interpretation
in terms of DM annihilation, with a cross section close to the thermal  value and with a DM mass around 50~GeV.
Recently, the GCE has also been confirmed by the {\it Fermi}-LAT collaboration \cite{TheFermi-LAT:2015kwa}.
In the present analysis, we will use the results from~\cite{Calore:2014xka}, where a detailed
spectral and morphological analysis of the excess has been performed and where the
inferred energy spectrum has been made available together with an error
covariance matrix.  The covariance includes an estimate of systematic uncertainties related to the galactic foreground emission, 
inferred from a grid of different foreground models and from a scan
of the typical model residuals along the galactic plane. 

\subsection{Annihilation cross section and photon spectrum}  \label{sec:ann}

\begin{figure}[htp]
\centering
\setlength{\unitlength}{1\textwidth}
\begin{picture}(0.91,0.2)
\put(0.02,-0.03){ 
 \put(-0.03,0.114){a)}
 \put(0.015,0.16){\small $S$}
 \put(0.015,0.06){\small $S$}
 \put(0.125,0.13){\small $h$}
 \put(0.23,0.16){\small SM}
 \put(0.23,0.06){\small SM}
 \put(0.04,0.06){\includegraphics[scale=0.5]{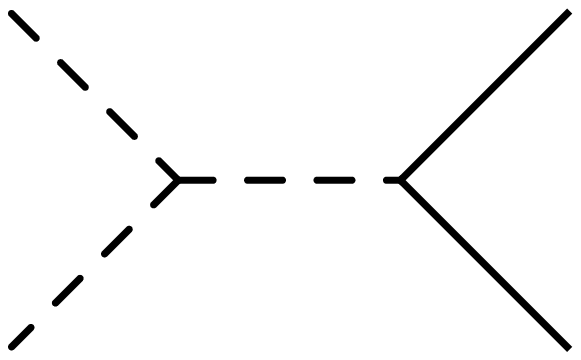}}
}
\put(0.39,-0.03){ 
 \put(-0.03,0.114){b)}
 \put(0.015,0.16){\small $S$}
 \put(0.015,0.06){\small $S$}
 \put(0.125,0.11){\small $S$}
 \put(0.2,0.16){\small $h$}
 \put(0.2,0.06){\small $h$}
 \put(0.04,0.06){\includegraphics[scale=0.5]{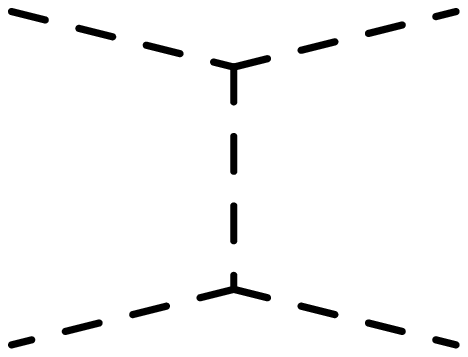}}
}
\put(0.74,-0.03){ 
 \put(-0.03,0.114){c)}
 \put(0.015,0.16){\small $S$}
 \put(0.015,0.06){\small $S$}
 \put(0.165,0.16){\small $h$}
 \put(0.165,0.06){\small $h$}

 \put(0.04,0.06){\includegraphics[scale=0.5]{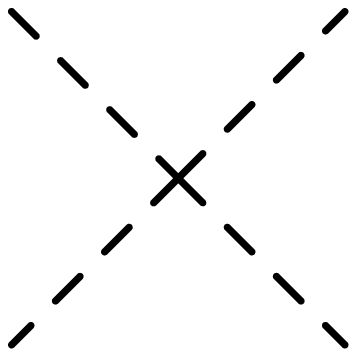}}
}
\end{picture}
\caption{Feynman diagrams for all WIMP annihilation processes. Below $m_S=m_h$ only
processes of type a) are present, where $\text{SM}=t,h,Z,W,b,\tau,c,g,\gamma$. 
Above the $hh$ threshold all three diagrams contribute.
}
\label{fig:annihilation_feyn}
\end{figure}

DM annihilation in the scalar Higgs portal model proceeds through $s$-channel Higgs and $t$-channel scalar exchange, and through the $S^2 h^2$ interaction, see figure~\ref{fig:annihilation_feyn}. 

Below $m_S = m_h$, only the $s$-channel Higgs diagram, figure~\ref{fig:annihilation_feyn} a), contributes to the annihilation cross section, and the relative strength of the different SM final states ($f = t,h,Z,W,b,\tau,c$, $g$, $\gamma$) is determined by the SM Higgs branching ratios, independent of the Higgs-scalar coupling $\lambda_{H\!S}$. 
Above the Higgs threshold, $m_S \ge m_h$, all diagrams depicted in figure~\ref{fig:annihilation_feyn} contribute, and, in particular, the $hh$ final state opens up. The strength of the annihilation into Higgs pairs, as compared to $W,Z$ or top-quark pairs, depends on the size of the Higgs-scalar coupling $\lambda_{H\!S}$. 

\begin{figure}[h!]
\centering
\setlength{\unitlength}{1\textwidth}
\begin{picture}(1,0.39)
 \put(-0.033,0.0){ 
  \put(0.07,0.03){\includegraphics[width=0.45\textwidth]{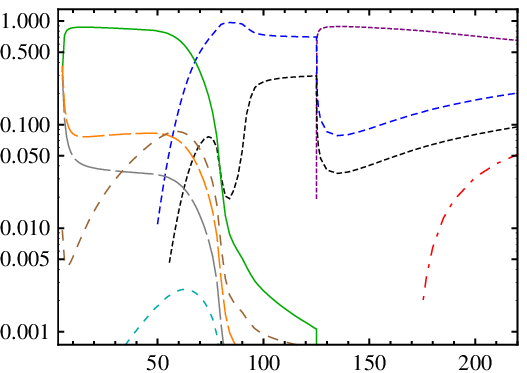}}
  \put(0.26,0.0){\footnotesize $m_S \,[\text{GeV}]$}
  \put(0.035,0.14){\rotatebox{90}{\footnotesize $\sigmav_f/\sigmav$}}
  \put(0.19,0.27){\rotatebox{0}{\tiny  \color{blue}  $WW$}}
  \put(0.285,0.24){\rotatebox{0}{\tiny   $ZZ$}}
  \put(0.18,0.09){\rotatebox{0}{\tiny  \color{ourcyan} $\gamma\gamma$ }}
  \put(0.42,0.145){\rotatebox{0}{\tiny \color{red}  $t \bar t $ }}
  \put(0.38,0.305){\rotatebox{0}{\tiny \color{ourpurple}  $hh $ }}
  \put(0.17,0.305){\rotatebox{0}{\tiny  \color{darkgreen} $b \bar b $ }}
  \put(0.133,0.193){\rotatebox{0}{\tiny \color{gray} $c \bar c $ }}
  \put(0.17,0.241){\rotatebox{0}{\tiny\color{orange}   $\tau \tau $ }}
  \put(0.144,0.145){\rotatebox{0}{\tiny \color{ourbrown}  $ gg $ }}
  \put(0.42,0.07){\rotatebox{0}{\scriptsize $\lambda_{H\!S}=1$ }}
  }
 \put(0.483,0.0){ 
  \put(0.07,0.03){\includegraphics[width=0.45\textwidth]{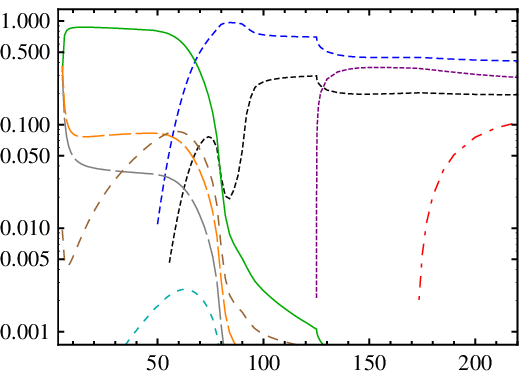}}
  \put(0.26,0.0){\footnotesize $m_S \,[\text{GeV}]$}
  \put(0.035,0.14){\rotatebox{90}{\footnotesize $\sigmav_f/\sigmav$}}
  \put(0.19,0.27){\rotatebox{0}{\tiny  \color{blue}  $WW$}}
  \put(0.285,0.24){\rotatebox{0}{\tiny   $ZZ$}}
  \put(0.18,0.09){\rotatebox{0}{\tiny  \color{ourcyan} $\gamma\gamma$ }}
  \put(0.413,0.15){\rotatebox{0}{\tiny \color{red}  $t \bar t $ }}
  \put(0.347,0.18){\rotatebox{0}{\tiny \color{ourpurple}  $hh $ }}
  \put(0.17,0.305){\rotatebox{0}{\tiny  \color{darkgreen} $b \bar b $ }}
  \put(0.133,0.193){\rotatebox{0}{\tiny \color{gray} $c \bar c $ }}
  \put(0.17,0.241){\rotatebox{0}{\tiny\color{orange}   $\tau \tau $ }}
  \put(0.144,0.145){\rotatebox{0}{\tiny \color{ourbrown}  $ gg $ }}
  \put(0.403,0.07){\rotatebox{0}{\scriptsize $\lambda_{H\!S}=0.01$ }}
  }
\end{picture}
\caption{Relative contribution to the dark matter annihilation cross section today, for two choices of the Higgs-scalar coupling $\lambda_{H\!S}=1$
(left panel) and $\lambda_{H\!S}=0.01$ (right panel).  Below $m_S=m_h$ the relative contribution is independent of $\lambda_{H\!S}$. 
}
\label{fig:anncontr}
\end{figure}

We have implemented the scalar Higgs portal model into \textsc{FeynRules}~\cite{Alloul:2013bka} and used \textsc{micrOMEGAs}~\cite{Belanger:2014vza}
linked to \textsc{CalcHEP}~\cite{Belyaev:2012qa} to compute the velocity-averaged annihilation cross sections $\sigmav_f$ for the various SM final states $f$. 
The loop-induced annihilation 
processes $SS\to gg, \gamma\gamma$ have been included using the effective Lagrangian \cite{Shifman:1979eb,Dawson:1990zj,Spira:1995rr}
 \begin{equation}
 {\cal L}_\text{HEFT} = \frac14 g^\text{eff}_{hgg} \,G^a_{\mu\nu}G^{\mu\nu,a} h +
 \frac14 g^\text{eff}_{h\gamma\gamma}\, F_{\mu\nu}F^{\mu\nu} h  
 \end{equation}
 where
\begin{align}
g^\text{eff}_{hgg} &= \frac{\alpha_\text{s}}{2 \pi v} \left(1+\frac{11\alpha_\text{s}}{4\pi}\right)\left| \sum_q A_{1/2}\!\left(\frac{4m_q^2}{s}\right)\right| \,,
\label{eq:geffhgg}\\
g^\text{eff}_{h\gamma\gamma} &= \frac{\alpha}{\pi v}
\left| \sum_q 3 Q_q^2 \,A_{1/2}\!\left(\frac{4m_q^2}{s}\right)+A_1\!\left(\frac{4m_W^2}{s}\right)\right| 
\label{eq:geffhaa}\,,
 \end{align}
and
\begin{align}
 A_{1/2}\left(\tau\right) & = \tau\left[1+(1-\tau)\arctan^2\left(\frac{1}{\sqrt{\tau-1}}\right)\right]\,,\\
A_{1}\left(\tau\right) & = -\frac12\left[2+3\tau +3(2\tau-\tau^2)\arctan^2\left(\frac{1}{\sqrt{\tau-1}}\right)\right]\,.
 \end{align}
Equation~\eqref{eq:geffhgg} takes into account the QCD corrections to the $ggh$ vertex~\cite{Dawson:1990zj,Spira:1995rr}. 
Here, $v=246\,$GeV is the vacuum expectation value of the Higgs, $\alpha_\text{s}$ and $\alpha$
are the strong and electromagnetic couplings, respectively, $s$ is the center-of-mass energy of the
process and $Q_q$ is the electric charge of the quark $q$. 
We take into account the contribution from the bottom and top quarks in the sum, $q=b,t$.
The strong coupling is evaluated at $s$, and we consider the one-loop running of $\alpha_\text{s}$.
We checked the accuracy of our implementation by comparing our results for the Higgs branching ratios as a function of the Higgs mass 
to those of the LHC Higgs Cross Section Working Group~\cite{Dittmaier:2011ti}. We find agreement within a relative
error below 5\% for $hgg$ and 15\% for $h\gamma\gamma$ in the mass region between 
$m_h=90$\,GeV (the minimal Higgs mass considered in~\cite{Dittmaier:2011ti}) and 
$m_h\simeq300$\,GeV (above which the respective contributions to the annihilation are 
completely irrelevant). Note that differences are expected as the results from Ref.~\cite{Dittmaier:2011ti} include 
further higher-order QCD and electroweak effects not taken into account in our calculation. For 
$m_S>m_h/2$ the total Higgs width in our model is identical to the SM Higgs width
and we take the theoretical prediction provided by the Higgs working group, 
$\Gamma_{\text{SM}}=4.03\,$MeV~\cite{Dittmaier:2011ti}. For $m_S\leq m_h/2$ we compute the invisible width 
$\Gamma_\text{inv}=\Gamma(h\to SS)$ first and run  \textsc{micrOMEGAs} with 
$\Gamma_h=\Gamma_\text{inv}+\Gamma_{\text{SM}}$ as an input parameter. We take the 
Higgs mass from the combined analysis of ATLAS and CMS~\cite{Aad:2015zhl}, $m_h=125.09$\,GeV.
The relative contributions of the different SM final states to the annihilation cross section is displayed 
in figure~\ref{fig:anncontr} for two different choices of the Higgs-scalar coupling, 
$\lambda_{H\!S} =  1$ and $\lambda_{H\!S} = 0.01$, respectively. 

The fragmentation, hadronization and decay of the SM particles from the primary annihilation process, figure~\ref{fig:annihilation_feyn}, produces a  spectrum of $\gamma$-rays, predominantly from $\pi^0\to\gamma\gamma$. The loop-induced process $S S \rightarrow h \rightarrow \gamma \gamma$ results in $\gamma$-ray lines at $E_\gamma \approx m_S$, and is suppressed as compared to the continuum photon spectrum from pion decays. Searches for $\gamma$-ray lines will be discussed in section~\ref{sec:constraints_indirect}. 
In general, $\gamma$-rays can also be produced from  electrons and positrons through inverse Compton scattering and synchrotron radiation. These contributions are important for DM particles annihilating predominantly into $e^+e^-$ and $\mu^+\mu^-$ final states \cite{Lacroix:2014eea}. In the scalar Higgs portal model, however, these annihilation channels are strongly suppressed, so that $\gamma$-rays from inverse Compton scattering and synchrotron radiation can be neglected in our analysis. 

We have generated the  $\gamma$-ray spectrum from SM particles, produced with a centre-of-mass energy $E = 2 m_S$, with the \textsc{Pythia~8.209} event generator~\cite{SjMrSk07}. The contributions of 3-body final states from annihilation into off-shell gauge bosons, $WW^*$ and $ZZ^*$, have been calculated using 
\textsc{Madgraph5\_aMC@NLO}~\cite{Alwall:2011uj,Alwall:2014hca}. Comparing the spectra calculated with the default \textsc{Pythia~8.209} generator to those obtained with \textsc{Pythia~6}~\cite{Sjostrand:2006za} we find differences of typically less than about 10\%, c.f.\ \cite{Cirelli:2010xx, Caron:2015wda}. Larger uncertainties occur for the $gg$ final state, which is however less significant for the overall $\gamma$-ray flux. We have compared the $\gamma$-ray spectra for the 2-body final states to those presented in Ref.~\cite{Cirelli:2010xx}. We find very good agreement in general, with some small deviations in the flux from annihilation into $gg$ and $c\bar{c}$ final states.   

The spectra from the various final states, $f = t,h,Z,W,b,\tau,c,g$, are combined, weighted by their relative strength as predicted within the scalar Higgs portal model as a function of the DM mass, $m_S$, and  the Higgs-scalar coupling, $\lambda_{H\!S}$, see figure~\ref{fig:anncontr}. The resulting $\gamma$-ray flux per unit solid angle at a photon energy $E_\gamma$ is 
\begin{equation}
 \frac{\textrm{d}\Phi}{\textrm{d}\Omega\textrm{d}E}   = \frac{1}{2 m_S^2} \sum_{f} \frac{\textrm{d}N_{f}}{\textrm{d}E} \frac{\langle \sigma v\rangle_{f}}{4 \pi} \int\limits_{\textrm{l.o.s}}\textrm{d}s\, \rho^2\left(r(s,\theta)\right)\,,
\end{equation}
where  $\textrm{d}{N}_f/\textrm{d}E$ is the photon spectrum per annihilation for a given final  state $f$, $\langle \sigma v\rangle_{f}$ is the corresponding velocity-averaged annihilation cross section, and $\rho$ is the DM density. The integral has to be evaluated  along the line-of-sight (l.o.s.) at an observational angle $\theta$ towards the galactic center. The l.o.s.\ integral of the DM 
density-squared over the solid angle $\textrm{d}\Omega$ is called the $J$-factor, and is discussed in more detail in section~\ref{sec:Jfactor}. 

\begin{figure}[h!]
\centering
\setlength{\unitlength}{1\textwidth}
\begin{picture}(0.62,0.35)
 \put(0.0,0.0){ 
  \put(0.0,-0.003){\includegraphics[width=0.6\textwidth]{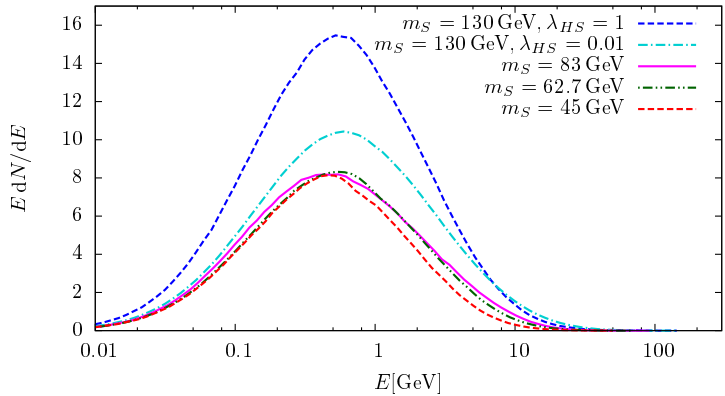}}
  }
\end{picture}
\caption{The scalar Higgs portal model prediction for the $\gamma$-ray spectrum per annihilation for different choices of $m_S$ and $\lambda_{H\!S}$. }
\label{fig:spectr}
\end{figure}

The scalar Higgs portal model prediction for the $\gamma$-ray spectrum per annihilation, $\sum_{f} \textrm{d}N_{f}/\textrm{d}E$ $\times \langle\sigma v\rangle_{f}/\langle \sigma v\rangle$, is shown in figure~\ref{fig:spectr} for different choices of $\lambda_{H\!S}$ and for dark matter masses $m_S$ that are of particular relevance in describing the GCE, as discussed below.

\subsection{Dark matter density profile and uncertainties}\label{sec:Jfactor}

The DM density in the Milky Way is only directly measured in the vicinity of the solar system, and only with a quite large
uncertainty, mostly systematic in nature. In the inner galaxy no direct measurements are available since the gravitational
potential is dominated by the baryonic matter.  Extrapolations are thus necessary together with assumptions about the
shape of the DM density profile,  which is typically parameterized as a cored or cuspy profile.
However, in the particular case of this analysis, where we are studying the DM interpretation of the GCE, we can 
limit the study to the DM profiles which are compatible with the measured shape of the GCE itself, i.e., cuspy profiles.

We will thus parameterize the DM profile as a generalized Navarro-Frenk-White (NFW) profile~\cite{Navarro:1995iw}:
\begin{equation}\label{eq:nfw}
\rho(r) =   \rho_s    \left(\frac{r}{r_s}\right)^{-\gamma}   \left(1+\frac{r}{r_s}\right)^{-3+\gamma},
\end{equation}
where $r$ is the spherical distance from the galactic center, 
and we will assume $\gamma=1.2\pm0.08$ (Gaussian error) as given in \cite{Calore:2014xka}.
For $\rho_s$ and $r_s$, the scale density and scale radius of the profile, respectively,
we will use the recent results from  \cite{Nesti:2013uwa} where the authors study 
a canonical NFW profile ($\gamma$=1) using up-to-date measurements of the rotation velocity of the Milky Way
as a function of the galacto-centric radius.  In particular they provide contours in the $\rho_s$-$r_s$ plane.
From the contours it can be seen that the single parameters are not well constrained, but they are tightly
correlated so that the contours can be simply approximated as a narrow line in the plane.
We found that the power law $\rho_s=42.7\,\textrm{GeV/cm}^3\cdot (r_s/\textrm{kpc})^{-1.59}$ 
describes well the relation among these two parameters. 
Formally, the above relation is valid only for $\gamma=1$ but we will use it for each $\gamma$ 
within the explored uncertainty of $\pm0.08$. This is expected to be a good approximation since
different values of $\gamma$  change the profile mainly in the inner kpcs from the galactic center, while the analysis
of  \cite{Nesti:2013uwa} is anyway performed for $r\gtrsim 2$ kpc, and is thus not very sensitive to moderate changes in $\gamma$.
We also note that the analysis of   \cite{Nesti:2013uwa} provides a  value of the DM local density $\rho_\odot = 0.471^{+0.048}_{-0.061}$ GeV cm$^{-3}$
which is thus implicit also in our analysis. This value has a relatively small error, which is typical of analyses based
on the Milky Way rotation curve (see also \cite{Catena:2009mf,Pato:2015dua,McMillan:2011wd}), 
while purely local analyses provide larger errors  $\rho_\odot \sim 0.2-0.7$ GeV cm$^{-3}$ \cite{Salucci:2010qr,Read:2014qva}.

\begin{figure}[h!]
\centering
\setlength{\unitlength}{1\textwidth}
\begin{picture}(0.6,0.355)
 \put(0.0,0.0){ 
  \put(0.03,0.0){\includegraphics[width=0.505\textwidth]{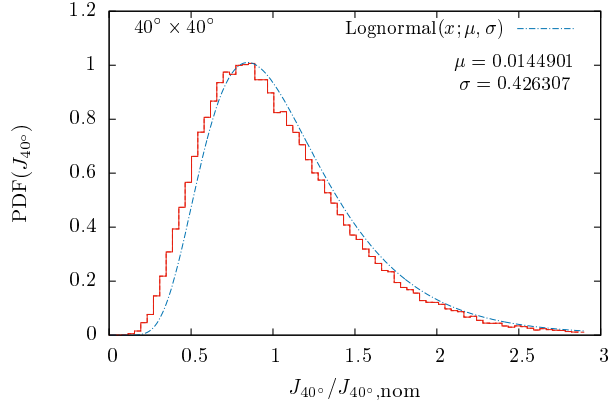}}
    }
\end{picture}
\caption{Probability density function of $J_{40^\circ}$, eq.~(\ref{eq:jfactor}). The red
histogram shows the distribution obtained from sampling the parameter $\gamma$
of the generalized NFW profile, eq.~(\ref{eq:nfw}), 
within its Gaussian uncertainty. The additional NFW parameters $\rho_s$ and $r_s$  are sampled uniformly, taking into account their correlation. The log-normal distribution fitted to the sampled distribution is shown in blue, with parameters $\mu$ and $\sigma$ as specified in the figure. }
\label{fig:Jbar1}
\end{figure}

Given the above $\rho_s$-$r_s$ relation and the further constraint $\gamma=1.2\pm0.08$, we 
determine the error on the GCE  $J$-factor with a Monte Carlo procedure.  The quantity of interest is 
the $J$-factor integrated over the sky region analysed in \cite{Calore:2014xka}, i.e., a $40^\circ \times 40^\circ$
region centred on the galactic center and with a stripe of $\pm 2^\circ$ masked along the galactic plane, 
\begin{equation}\label{eq:jfactor}
J_{40^\circ} =  \int\limits_{\Delta \Omega}\!\textrm{d}\Omega \! \int\limits_{\textrm{l.o.s}}\!\textrm{d}s \rho^2{\left(r(s,\theta)\right)}\,.
\end{equation}
In figure~\ref{fig:Jbar1} we show the distribution of $J_{40^\circ}$ that we obtain from sampling $\gamma$
within its Gaussian uncertainty and with $\rho_s$ and $r_s$ uniformly distributed, taking into account their correlation.
It can be seen that  $J_{40^\circ}$ is well approximated by a log-normal distribution with 
a width of $\sigJ\simeq 0.43$ (see figure~\ref{fig:Jbar1}).
In the following we will use this $J_{40^\circ}$ distribution   to account for its uncertainty.
$J_{40^\circ\!,\,\text{nom}}$ is the nominal value of $J_{40^\circ}$ for $\gamma=1.2$  $\rho_s=0.74\,\textrm{GeV/cm}^3$, $r_s=19.5\,\textrm{kpc}$, 
i.e., $J_{40^\circ\!,\,\text{nom}}= 1.79\cdot 10^{23}\,\textrm{GeV}^2 \textrm{cm}^{-5}$.
We also use $r_\odot=8.0$ kpc, although we verified that  $J_{40^\circ}$ varies very little
varying $r_\odot$ in the range 7.5-8.5 kpc, which is the typical uncertainty on $r_\odot$.
Note that since the authors of \cite{Calore:2014xka} normalise the GCE flux dividing by the angular size of the analysed region,
we also need to divide $J_{40^\circ}$ by the corresponding  solid angle $\Delta\Omega=0.43$ sr.

\subsection{WIMP contribution to dark matter}\label{sec:WIMP}

In this study we allow for the possibility that the dark sector is more complex than 
containing just one DM particle species. We could, for example, imagine a second non-WIMP DM 
component (such as axions or primordial black holes) which does not interact weakly with the SM and which does not annihilate into SM particles today.
Hence, we consider the case that the WIMP DM density is a certain fraction, $R\leq1$, of
the total (gravitationally interacting) DM:
\begin{equation}
\rho_\text{WIMP} = R \,\rho_{\text{total}}\,.
\end{equation}
Here we assume that there is no difference in the clustering properties and hence 
in the density profiles of the WIMP and non-WIMP DM components.
The annihilation signal today thus scales as $\phi\propto R^2$. We will consider $R$ as 
a free parameter in the fit of the GCE signal. 

\subsection{Fit to the GCE signal}\label{sec:gcefit}

In order to perform a fit to the GCE we use \textsc{MultiNest}~\cite{Feroz:2008xx,Feroz:2013hea},
which allows to scan the parameter space under study much more efficiently than a simple random search.
The respective parameters and scan ranges are summarized in table~\ref{tab:scanparams}.
The annihilation cross sections are computed using \textsc{micrOMEGAs}. The $\chi^2$ 
for the GCE (including the contribution from $J_{40^\circ}$) is computed as: 
\begin{equation}
\label{eq:chi2GCE}
\chi_\text{GCE}^2 = \sum_{i,j} (d_i - t_i ) \left( \Sigma_{ij} + \delta_{ij} (\sigma_{\text{rel}}\ t_i)^2\right)^{-1}(d_j -  t_j) 
 + \frac{(\log J_{40^\circ} - \log J_{40^\circ\!,\,\text{nom}} )^2}{(\sigJ)^2},
\end{equation}
where $d_i$ is  the  GCE measured flux in energy bin $i$ from \cite{Calore:2014xka},
$t_i$ is our model prediction, which depends on the parameters $m_S$, $\lamhs$, $R$ and $J_{40^\circ}$,
$\Sigma_{ij}$ is the covariance matrix given in   \cite{Calore:2014xka}, which includes statistical and systematic errors,
and  $ J_{40^\circ\!,\,\text{nom}}$ and $\sigJ$ are as defined in section~\ref{sec:Jfactor}.
The term  $ \delta_{ij} (\sigma_{\text{rel}}\ t_i)^2$ represents   a diagonal error equal to a fraction $\sigma_{\text{rel}}$
of the model prediction itself, which we add to the original $\Sigma_{ij}$  in order to take into account the model uncertainties in the annihilation spectrum.
We choose $\sigma_{\text{rel}}$ =  10\%, as discussed in section~\ref{sec:ann}.
 Since we include in the error a dependence from the model itself, the likelihood 
which we use in the fit reads slightly differently from the $\chi^2$ expression in eq.~(\ref{eq:chi2GCE}).
Specifically, up to a constant factor which has no influence on the fit, the log-likelihood is
\begin{equation}
\label{eq:logLGCE}
-2 \log \mathcal{L}_\text{GCE}   =     \chi_\text{GCE}^2   + \log | \Sigma_{ij} + \delta_{ij} (\sigma_{\text{rel}}\ t_i)^2 |,
\end{equation}
where $| \Sigma_{ij} + \delta_{ij} (\sigma_{\text{rel}}\ t_i)^2 |$ is the determinant of the covariance matrix.

\begin{table}[t]
\begin{center}
\renewcommand{\arraystretch}{1.2}
\begin{tabular}{c | c} 
parameter & range  \\ \hline
$m_S$ & $[5;220]\,\text{GeV}$\\
$\lambda_{H\!S}$  & $[3\times10^{-5};4\pi]$\\ \hline
$\logJ$& $[-4\sigJ ; 4\sigJ]$\\
$R$ & $[10^{-3};1]$\\
\end{tabular}
\renewcommand{\arraystretch}{1}
\end{center}
\caption{Fit parameters and their corresponding ranges. For the special case of $R=1$ and 
$J_{40^\circ} = J_{40^\circ\!,\,\text{nom}}$  the ranges for $m_S$ and $\lambda_{H\!S}$ are the same.}
\label{tab:scanparams}
\end{table}

\begin{figure}[h!]
\centering
\setlength{\unitlength}{1\textwidth}
\begin{picture}(0.65,0.65)
 \put(0.0,0.0){ 
   \put(0.01,0.01){\includegraphics[width=0.70\textwidth]{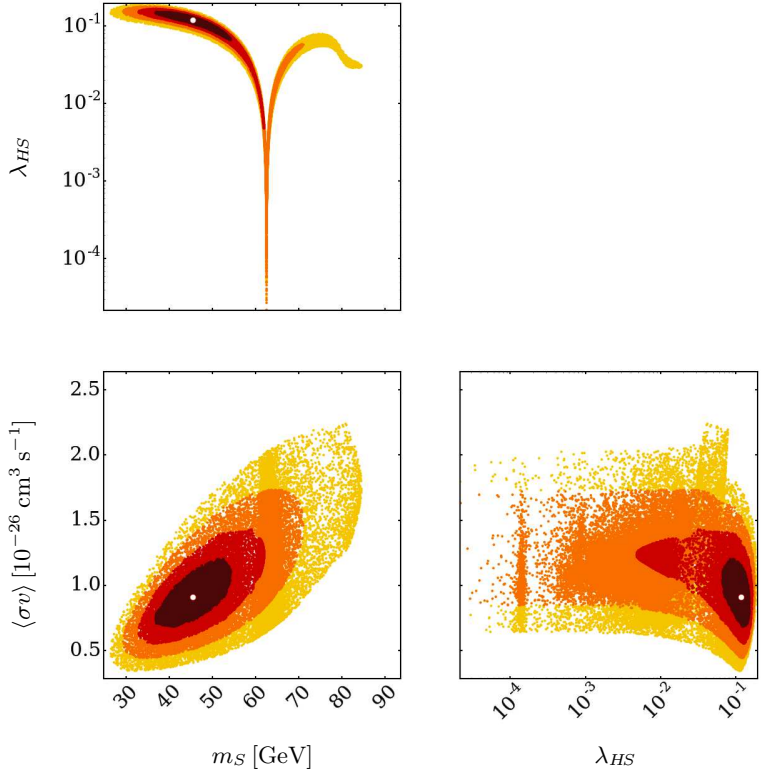}}
  \put(0.187,0.01){\footnotesize $m_S \,[\text{GeV}]$}
  \put(0.505,0.01){\footnotesize $\lamhs$}
  \put(0.019,0.12){\rotatebox{90}{\footnotesize $\sigmav \,[10^{-26}\,\text{cm}^3\,\text{s}^{-1}]$}}
  \put(0.019,0.495){\rotatebox{90}{\footnotesize $\lamhs$}}
    }
\end{picture}
\caption{Results of a fit to the GCE with free parameters $m_S$ and $\lambda_{H\!S}$ and with $R=1$ and $ \logJ=0$ fixed. 
We also show the annihilation
cross section today, $\sigmav$. 
The white dot denotes the best-fit point. The dark-red, red, orange and yellow points lie within the
1, 2, 3 and $4\sigma$ region around the best-fit point, respectively. We take into account the
log-likelihood from the GCE only.
}
\label{fig:Req1gc}
\end{figure}

\textsc{MultiNest} is particularly suited for Bayesian analyses,
since it naturally provides a sample of the posterior distribution, i.e., the product of
the likelihood times the  priors for the parameters. Nonetheless, 
the results of the scan of the parameter
space from \textsc{MultiNest} can also be  used in the frequentist framework, provided that the  posterior, and in turn the likelihood,
has been explored in enough detail.
The advantage of the frequentist interpretation is that the derived constraints are not dependent on
the prior chosen to explore the various parameters.  We will thus  adopt the frequentist interpretation in the following. 
Within this framework marginalisation over parameters is performed with the profile likelihood method \cite{Rolke:2004mj}
and contours at a certain confidence level are drawn following the expectation of a $\chi^2$ distribution.
For example, for contour plots in two dimensions  we first derive  the profile
likelihood in the two given parameters  profiling over the remaining ones. We then draw contours 
around the best-fit at 1, 2, 3 and 4$\sigma$ confidence
level  according to a two-dimensional $\chi^2$ distribution.
A further advantage of using the frequentist formalism is that the output of 
different \textsc{MultiNest} scans can be easily
combined, as we indeed do in figure~\ref{fig:Req1gc} (and subsequent figures) where different densites of points
arise from different separate scans. 
A disadvantage of this procedure is, of course, that the density of points, which in the Bayesian interpretation has
a precise meaning (i.e., it traces the posterior distribution)
now loses any meaning (i.e., in figure~\ref{fig:Req1gc} and subsequent figures only the color of the points is important, not the density).
To ensure that the likelihood is well sampled,  we run \textsc{MultiNest} with high-accuracy settings,
using between 1000 and 3000 live points, depending on the scan, a typical tolerance $\texttt{tol}=$ 0.001, and an enlargement factor between $\texttt{efr}=0.3-0.5$ in order
to ensure that also the tails of the distribution are well explored. 

Previous analyses of the GCE within the scalar Higgs portal model have considered a simple  
dark sector with a scalar WIMP particle that constitutes all of DM, and no uncertainty in the DM density profile. 
In our analysis, such a simplified scenario corresponds to the special case where $R= \rho_\text{WIMP}/\rho_{\text{total}} = 1$ and $J_{40^\circ}= J_{40^\circ\!,\,\text{nom}}$. 
Figure~\ref{fig:Req1gc} shows the results of the corresponding fit of the model parameters $m_S$ and $\lamhs$, with $R=1$ and $J_{40^\circ}= J_{40^\circ\!,\,\text{nom}}$ fixed. We also 
show the derived parameter $\sigmav$. A good fit is only achieved in a narrow Y-shaped region for which $\sigmav$ is of the order of
$10^{-26}\,\text{cm}^3/\text{s}$. Figure~\ref{fig:Req1gc} also shows that the annihilation cross section $\sigmav$ has to increase with increasing $m_S$ (see lower left panel) as to provide an 
overall $\gamma$-ray flux consistent with the GCE, compensating the reduced DM density which decreases as $1/m_S^2$.

On the other hand, as discussed in the previous section, the dark sector may well be more complex than containing just one WIMP DM species. We thus allow 
$R= \rho_\text{WIMP}/\rho_{\text{total}} \le 1$ and consider $R$ an input parameter for our fit. Moreover, we include the 
astrophysical uncertainties in the $J$-factor as explained above. Our full fit of the GCE within the Higgs scalar portal model thus 
contains the parameters of the model, $m_S$ and $\lambda_{H\!S}$, and two additional parameters related to the astrophysical 
scenario, $R$ and $\logJ$.

\begin{figure}[h!]
\centering
\setlength{\unitlength}{1\textwidth}
\begin{picture}(1,1.01)
 \put(0.0,0.0){ 
  \put(-0.022,-0.02){\includegraphics[width=1.1\textwidth]{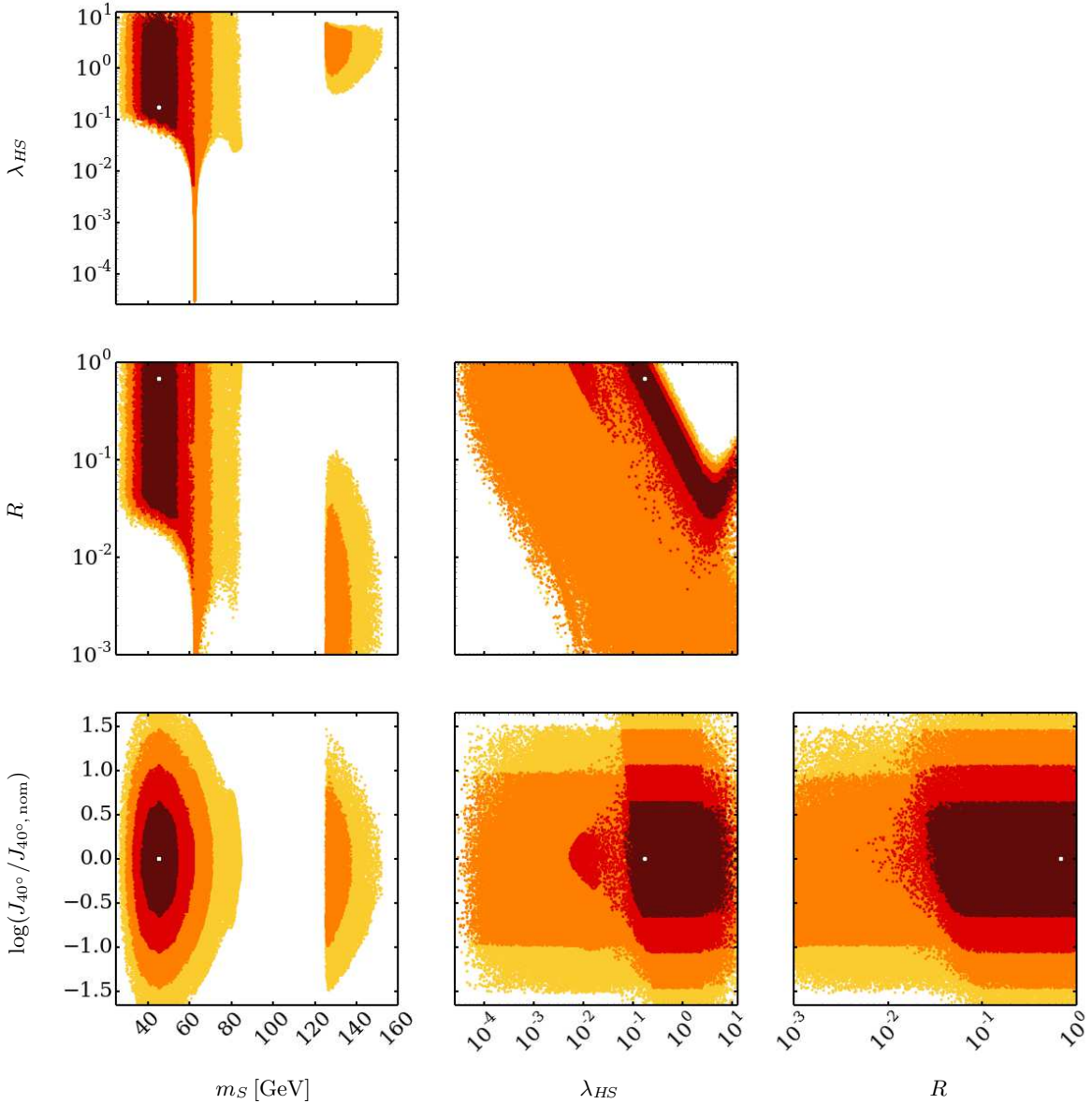}}
  \put(0.204,0.01){\footnotesize $m_S \,[\text{GeV}]$}
  \put(0.527,0.01){\footnotesize $\lamhs$}
  \put(0.838,0.01){\footnotesize $R$}
  \put(0.019,0.135){\rotatebox{90}{\footnotesize $\logJ$}}
  \put(0.019,0.52){\rotatebox{90}{\footnotesize $R$}}
  \put(0.019,0.824){\rotatebox{90}{\footnotesize $\lamhs$}}
  }
\end{picture}
\caption{Results of a fit to the GCE  with free parameters $m_S$, $\lambda_{H\!S}$, $R$ and 
$\logJ$. 
The white dot denotes the best-fit point. The dark-red, red, orange and yellow points lie within the
1, 2, 3 and $4\sigma$ region around the best-fit point, respectively. We take into account the
log-likelihood from the GCE only.
}
\label{fig:Rneq1gc}
\end{figure}

Figure~\ref{fig:Rneq1gc} shows the result of the complete fit of the GCE for the four input parameters 
$m_S$, $\lambda_{H\!S}$, $R$ and $\logJ$. Instead of the narrow Y-shaped stripe with $\lambda_{H\!S} \lesssim 0.1$ which we observe 
for the case $R=1$ (figure~\ref{fig:Req1gc}), the region of larger $\lamhs$ is now also allowed  as $R<1$ can compensate for the larger annihilation cross section (see upper left panel).   
In addition, as $R\neq1$ relaxes the tight connection between the normalization
(governed by $\sigmav R^2$) and the shape 
of the photon spectrum (governed by $\lamhs$ for a given mass $m_S$), a new region appears for DM masses above the Higgs threshold, $m_S\geq m_h$.
At the $hh$-threshold the spectral shape for annihilation into $hh$ fits better than for $WW$
and so large $\lamhs$ are preferred. However, for large $\lamhs$ the overall $\gamma$-ray flux is too large to accommodate the GCE signal unless
$R$ is small. Hence, for $m_S>m_h$ the fit prefers small $R$, see middle left panel in figure~\ref{fig:Rneq1gc}. 

\section{Constraints}\label{sec:constraints}

The scalar singlet Higgs portal model is constrained by
the Higgs invisible width, by direct detection searches, by searches for $\gamma$-rays from dwarf spheroidal galaxies, by searches for $\gamma$ spectral lines, and by the DM relic density.  
We will discuss these constraints in turn below and quantify their impact by including the corresponding likelihoods into the global fit of the GCE signal. 
We have checked that constraints for CMB anisotropies \cite{Slatyer:2015jla} on the annihilation cross section are less constraining than limits from dwarf spheroidal galaxies in the 
relevant region of parameter space (10\,GeV $\lesssim m_S\lesssim 1\,$TeV). 
Limits on the scalar Higgs portal model from CMB anisotropies have been considered in \cite{Cline:2013gha, Beniwal:2015sdl}.

\subsection{Higgs invisible width} \label{sec:BRinv}
For light scalar DM particles below the Higgs threshold, $m_S < m_h/2$, the decay $h \to SS$ results in an invisible Higgs width, $\Gamma_{\rm inv}$. 
This region of parameter space is thus constrained by the LHC limits on the Higgs invisible branching ratio, ${\rm BR}_{\rm inv} \lesssim 0.23$~\cite{Aad:2015pla}.

In the scalar Higgs portal model, the invisible Higgs width is \cite{Kanemura:2010sh}
\begin{equation}
\Gamma_{\rm inv} = \Gamma(h\to SS) = \frac{\lambda_{H\!S}^2 v^2}{32 \pi m_h}\sqrt{1-\frac{4m_S^2}{m_h^2}}\,.
\end{equation}
Assuming that the visible Higgs decay width is given by the Standard Model width, $\Gamma_{\rm SM}$, so that ${\rm BR}_{\rm inv} = \Gamma_{\rm inv}/(\Gamma_{\rm SM} + \Gamma_{\rm inv})$, the upper limit on ${\rm BR}_{\rm inv}$ 
implies an upper limit on $\Gamma_{\rm inv}$ and thus on the Higgs-scalar coupling $\lambda_{H\!S}$ as a function of the DM mass. For the numerical analysis we have used the value 
$\Gamma_{\text{SM}}=4.03\,$MeV~\cite{Dittmaier:2011ti}
for the Standard Model Higgs width and the log-likelihood function for ${\rm BR}_{\rm inv}$ provided by the ATLAS analysis~\cite{Aad:2015pla}. 
     
\subsection{Direct detection}  \label{sec:LUX}
A spin-independent DM-nucleon scattering cross section is predicted within the scalar Higgs portal model through the exchange of the SM Higgs boson. The model 
is therefore severely constrained by direct detection experiments. As the mass of the SM Higgs is large compared to the momentum transfer in the elastic DM-nucleon scattering, 
the cross section can be described by an effective interaction and is given by~\cite{Cline:2013gha}
\begin{equation} 
\label{eq:sigmaSI}
\sigma_\text{SI} = \frac{\lambda_{H\!S}^2 f_\text{N}^2}{4 \pi}\frac{\mu_\text{r}^2 m_{\rm N}^2}{m_h^4 m_S^2}\,.
\end{equation}
Here, $f_\text{N} = 0.30$~\cite{Cline:2013gha} denotes the strength of the effective Higgs-nucleon interaction, and $\mu_{\rm r} = m_{\rm N} m_S/(m_{\rm N}+m_S)$ is the DM-nucleon reduced mass. The current best limits on $\sigma_\text{SI}$ come from the LUX experiment~\cite{Akerib:2013tjd}. To obtain the likelihood and $p$-value of the LUX direct detection limits, we have used the tool LUXCalc~\cite{Savage:2015xta}, where the likelihood is constructed from a Poisson distribution. For more details we refer to Ref.~\cite{Savage:2015xta}. In section~\ref{sec:results} we shall also comment on the projected sensitivity of future direct detection experiments like XENON1T~\cite{Aprile:2015uzo} or DARWIN~\cite{Baudis:2012bc}. 
Note that in contrast to indirect detection, a direct detection signal scales 
linearly with the DM density and hence linearly with $R$.

\subsection{Indirect detection: dwarf spheroidal galaxies and spectral $\gamma$ lines} \label{sec:constraints_indirect}

The GCE can be tested using other independent $\gamma$-ray observations and analyses. At present, observations of dwarf satellite galaxies of the Milky Way provide the most stringent limits
on $\langle \sigma v \rangle$. There is indeed a mild tension between the DM interpretation
of the GCE and dwarf limits \cite{Ackermann:2015zua}. However, taking into account the respective uncertainties of the $J$-factors of the galactic center and of the dwarf galaxies, 
the GCE signal can be accommodated, as we shall quantify below. 

To implement the dwarf constraints we use the tabulated likelihood as function of flux for each dwarf provided in \cite{Ackermann:2015zua}.
We write the likelihood as a product of likelihoods over each single dwarf as described in \cite{Ackermann:2015zua} and \cite{Ahnen:2016qkx}. 
In particular we consider the seven most constraining dwarfs:  Willman 1, Ursa Minor, Ursa Major II,  Segue 1, Draco, Coma Berenices and Bootes I.
The likelihood of each dwarf contains a factor which depends on $\langle \sigma v \rangle$ and the provided tabulated flux likelihood,
and a further log-normal factor describing the uncertainty in the $J$-factor, $J_i$, of the dwarf. 
For the latter, we use the nominal $J_i$ and uncertainty provided in \cite{Ackermann:2015zua,Ahnen:2016qkx}.
The seven $J_i$ of the dwarfs are profiled during our global fit, i.e., for each point sampled in parameter space
we tabulate \emph{on-the-fly} the dwarf's likelihood as function of $J_i$ and take the corresponding maximum likelihood value. 

We also use the results from the latest search for $\gamma$-ray lines in the inner galaxy \cite{Ackermann:2015lka},
which set constraints on the  annihilation cross section into mono-chromatic photons, $\langle \sigma v \rangle_{\gamma\gamma}$. 
In  \cite{Ackermann:2015lka} different regions of interest (ROI) are analysed, each maximising the sensitivity to a possible DM
annihilation signal for different DM density profiles.
We use the results for both the R3 and R16 ROI, corresponding to a circular region of 3$^{\circ}$ and 16$^{\circ}$ of radius, respectively, see \cite{Ackermann:2015lka}. 
These ROIs maximise the sensitivity for a generalised NFW profile with an inner slope $\gamma=1.3$ and
for an Einasto profile which is slightly more cored, respectively.
To derive the  likelihood  from searches for $\gamma$-ray lines as function of the integrated flux in the given ROI,
we take from  \cite{Ackermann:2015lka}  for each energy the fluxes corresponding to the 95\% upper limit  ($\Delta\log L=2.71/2$), to $\Delta\log L=1.0/2$ and
to $\Delta\log L=0.0.$\footnote{Andrea Albert, private communication.} We then approximate the log-likelihood for each energy
as a parabola passing through the above points when the minimum is at a positive flux, or as a line when
the minimum is at zero flux. 

\begin{figure}[h!]
\centering
\setlength{\unitlength}{1\textwidth}
\begin{picture}(0.5,0.354)
 \put(0.0,0.0){ %
  \put(0.0,0.02){\includegraphics[width=0.5\textwidth]{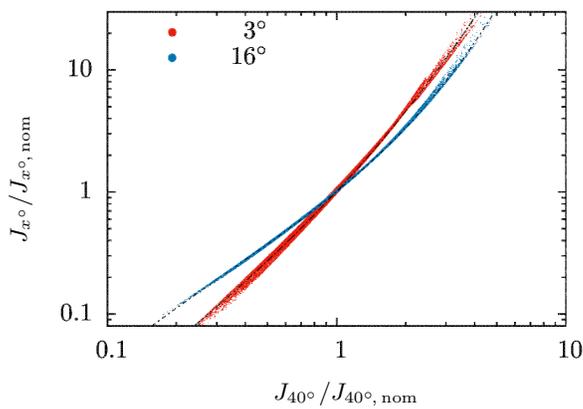}}
  \put(0.22,0.0){\footnotesize $J_{40^\circ}/J_{40^\circ\!,\,\text{nom}}$}
  \put(0.0,0.13){\rotatebox{90}{\footnotesize $J_{x^\circ}/J_{x^\circ\!,\,\text{nom}}$}}
  }
\end{picture}
\caption{Correlation between $J_{40^\circ}$ and 
the $J$-factors of the two ROIs relevant for the $\gamma$ line searches. The correlations are obtained from sampling the parameter $\gamma$
of the generalized NFW profile, eq.~(\ref{eq:nfw}), within its Gaussian uncertainty.  The additional NFW parameters $\rho_s$ and $r_s$  are sampled uniformly, taking into account their correlation. The red and blue
points denote $J_{3^\circ}$ (steeper behaviour) and $J_{16^\circ}$,
respectively. We also show the fit quartic polynomials in log-log space
(dot-dashed lines).
}
\label{fig:Jbar2}
\end{figure}

To translate the above likelihood for the flux into a likelihood for the annihilation cross section $\langle \sigma v \rangle_{\gamma\gamma}$, the $J$-factors corresponding to the ROIs ${J}_{3^\circ}$ and ${J}_{16^\circ}$ are needed, see eq.~(\ref{eq:jfactor}). Using the Monte Carlo procedure described in section \ref{sec:Jfactor}
we find, not surprisingly, a strong correlation between $J_{40^\circ}$ and $J_{3^{\circ}},J_{16^\circ}$,
which is shown in figure~\ref{fig:Jbar2}. This correlation is well
approximated by a forth order polynomial in log-log space, also shown in figure~\ref{fig:Jbar2}.
Thus, instead of considering $J_{3^{\circ}}$ and $J_{16^\circ}$ as further independent nuisance parameters of the fit, we only use $J_{40^\circ}$
as free parameter, with the log-normal distribution described in section~\ref{sec:Jfactor} accounting for the uncertainty related to the DM profile.
$J_{3^\circ}$ and $J_{16^\circ}$ are considered functions of $J_{40^\circ}$ with no further intrinsic uncertainties. 

We calculate  $\langle \sigma v \rangle_{\gamma\gamma}$ using the Higgs effective Lagrangian as described in section~\ref{sec:ann}.

\subsection{Relic density}  \label{sec:reldens}

Assuming a standard cosmological history, we can link the 
relic WIMP density from the thermal freeze-out to the DM density as measured 
by Planck, $\Omega h^2|_\text{Planck} = 0.1198\pm 0.0015$~\cite{Ade:2015xua}. 
The total DM density predicted by our model is
\begin{equation}
\Omega h^2|_\text{DM,\,total}=\frac{\Omega h^2|_\text{WIMP}}{R}\,.
\end{equation}
We compute $\Omega h^2|_\text{WIMP}$ with \textsc{micrOMEGAs}.
For details regarding the model implementation we refer to section~\ref{sec:ann}. We include the effective Higgs-gluon coupling
$g^\text{eff}_{hgg}$, eq.~(\ref{eq:geffhgg}), which depends on the center-of-mass energy of the annihilation process, $s$. In contrast to the case of annihilation today, for the computation of the 
annihilation cross section during freeze-out $\sqrt{s}=2m_S$ is not always a good approximation:
for $2m_S$ just below $m_h$, the dominant contribution to the thermally averaged 
cross section $\sigmav$ comes from the resonance, $\sqrt{s}=m_h$. 
Hence, for the computation of $g^\text{eff}_{hgg}$, eq.~(\ref{eq:geffhgg}), we choose $\sqrt{s}=m_h$ for $m_S<m_h/2$.

We compute the $\chi^2$ for the relic density constraint from
\begin{equation}
\chi_\Omega^2 = \frac{\left(\Omega h^2|_\text{DM,\,total}-\Omega h^2|_\text{Planck}\right)^2}{\left(\sigma_\text{rel}\times\Omega h^2|_\text{DM,\,total}\right)^2}\,,
\end{equation}
where we assume that the dominant uncertainty comes from the theoretical prediction of the
relic density, $\sigma_\text{rel}=10\%$. The respective log-likelihood reads
\begin{equation}
-2 \log \mathcal{L}_\Omega   =     \chi_\Omega^2   
+ 2 \log (\sigma_{\text{rel}}\,\Omega h^2|_\text{DM,\,total}),
\end{equation}
again, up to an irrelevant constant.

\section{Results and discussion} \label{sec:results} 

In this section we shall present the results of a global fit to the GCE, taking into account the constraints from the invisible Higgs branching ratio, direct detection limits, independent searches for $\gamma$-rays from dwarf satellite galaxies, searches for spectral $\gamma$ lines, and from the dark matter relic density, as discussed in section~\ref{sec:constraints}.

In section~\ref{sec:gcefit} we have shown that the GCE signal can be well described by the scalar Higgs portal model. We found that a small DM mass, $30~{\rm GeV} \lesssim m_S \lesssim 100~{\rm GeV}$, provides the best fit, and that small values of the Higgs-scalar coupling, $\lambda_{H\!S} \gtrsim 10^{-4}$,  are viable near the Higgs resonance $m_S \approx m_h/2$. Parameter regions above the Higgs threshold, $m_S \gtrsim m_h$ are viable, too, but only if we allow for a significant non-WIMP contribution to DM, corresponding to small values of $R= \rho_\text{WIMP}/\rho_{\text{total}}$. For $m_S \gtrsim m_h$ and large $\lambda_{H\!S}$ of ${\cal O}(1)$, the $hh$ final state is dominant and provides a good description of the shape of the $\gamma$-ray spectrum. Small values of $R$ are required in this parameter region to reconcile the corresponding large annihilation cross section $\sigmav$ with the GCE flux $\propto \sigmav \times R^2$.

\begin{figure}[h!p]
\centering
\setlength{\unitlength}{1\textwidth}
\begin{picture}(1,1.1)
 \put(-0.03,0.0){ 
  \put(0.0,0.03){\includegraphics[width=0.55\textwidth]{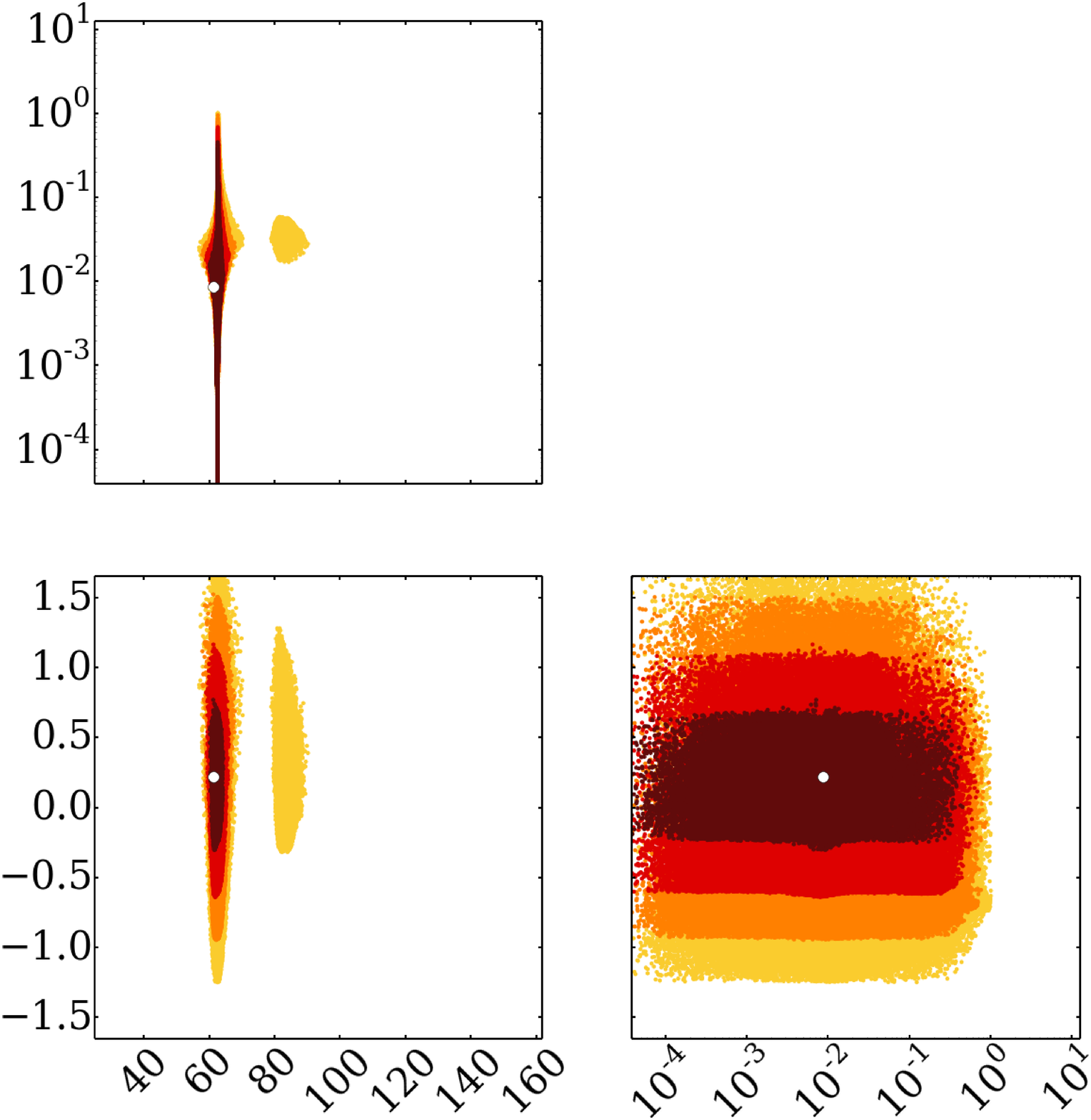}}
  \put(0.157,0.555){\footnotesize \underline{GCE+$\text{BR}_\text{inv}$+LUX+dwarfs}}
  \put(0.127,0.01){\footnotesize $m_S \,[\text{GeV}]$}
  \put(0.39,0.01){\footnotesize  $\lamhs$}
  \put(-0.01,0.105){\rotatebox{90}{\footnotesize $\logJ$}}
  \put(-0.01,0.4){\rotatebox{90}{\footnotesize  $\lamhs$}}
  }
 \put(0.52,0.0){ 
  \put(0.0,0.03){\includegraphics[width=0.55\textwidth]{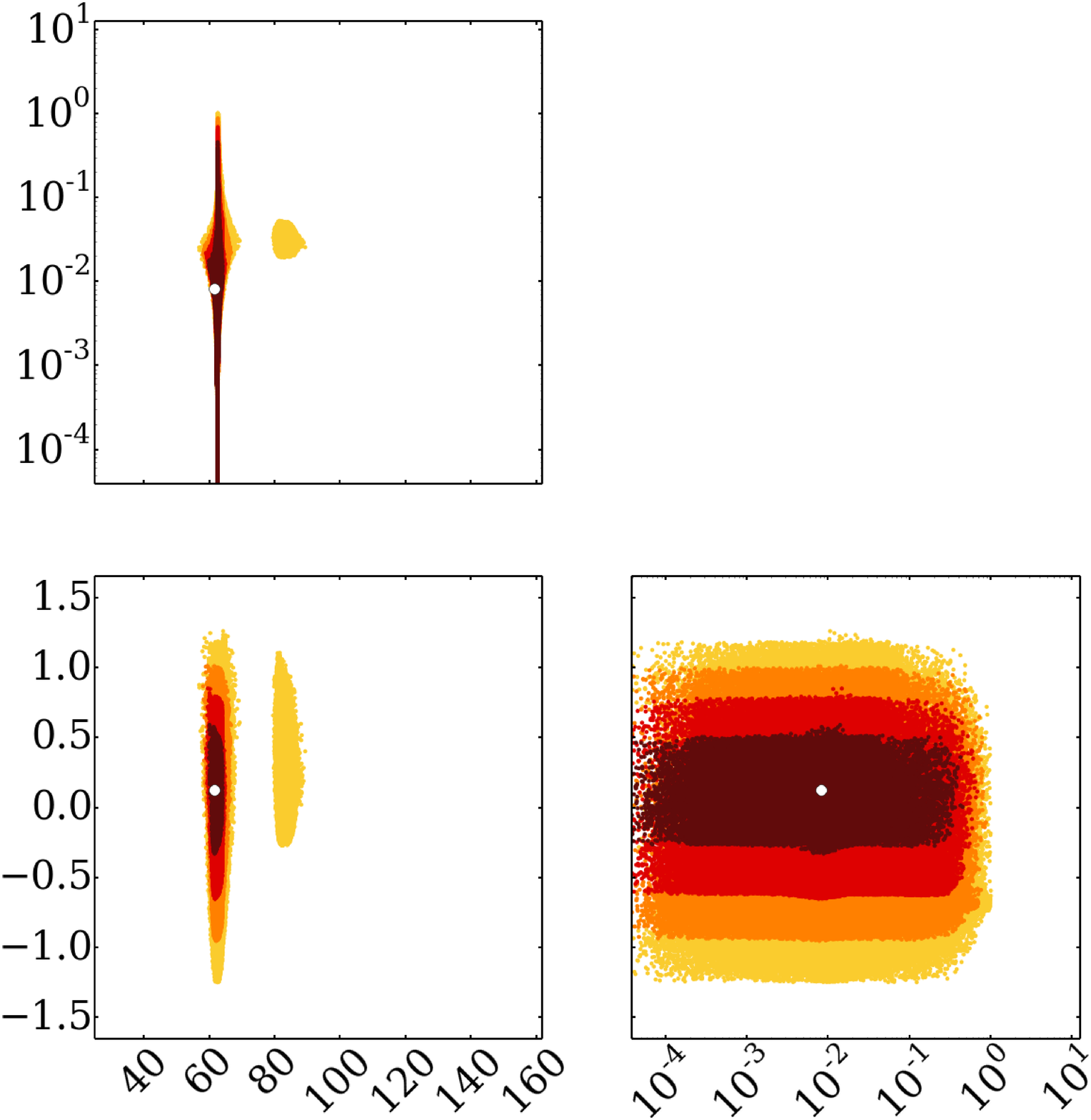}}
  \put(0.133,0.555){\footnotesize \underline{GCE+$\text{BR}_\text{inv}$+LUX+dwarfs+lines}}
  \put(0.127,0.01){\footnotesize $m_S \,[\text{GeV}]$}
  \put(0.39,0.01){\footnotesize  $\lamhs$}
  \put(-0.01,0.105){\rotatebox{90}{\footnotesize $\logJ$}}
  \put(-0.01,0.4){\rotatebox{90}{\footnotesize  $\lamhs$}}
  }
 \put(-0.03,0.62){ 
  \put(0.0,0.03){\includegraphics[width=0.55\textwidth]{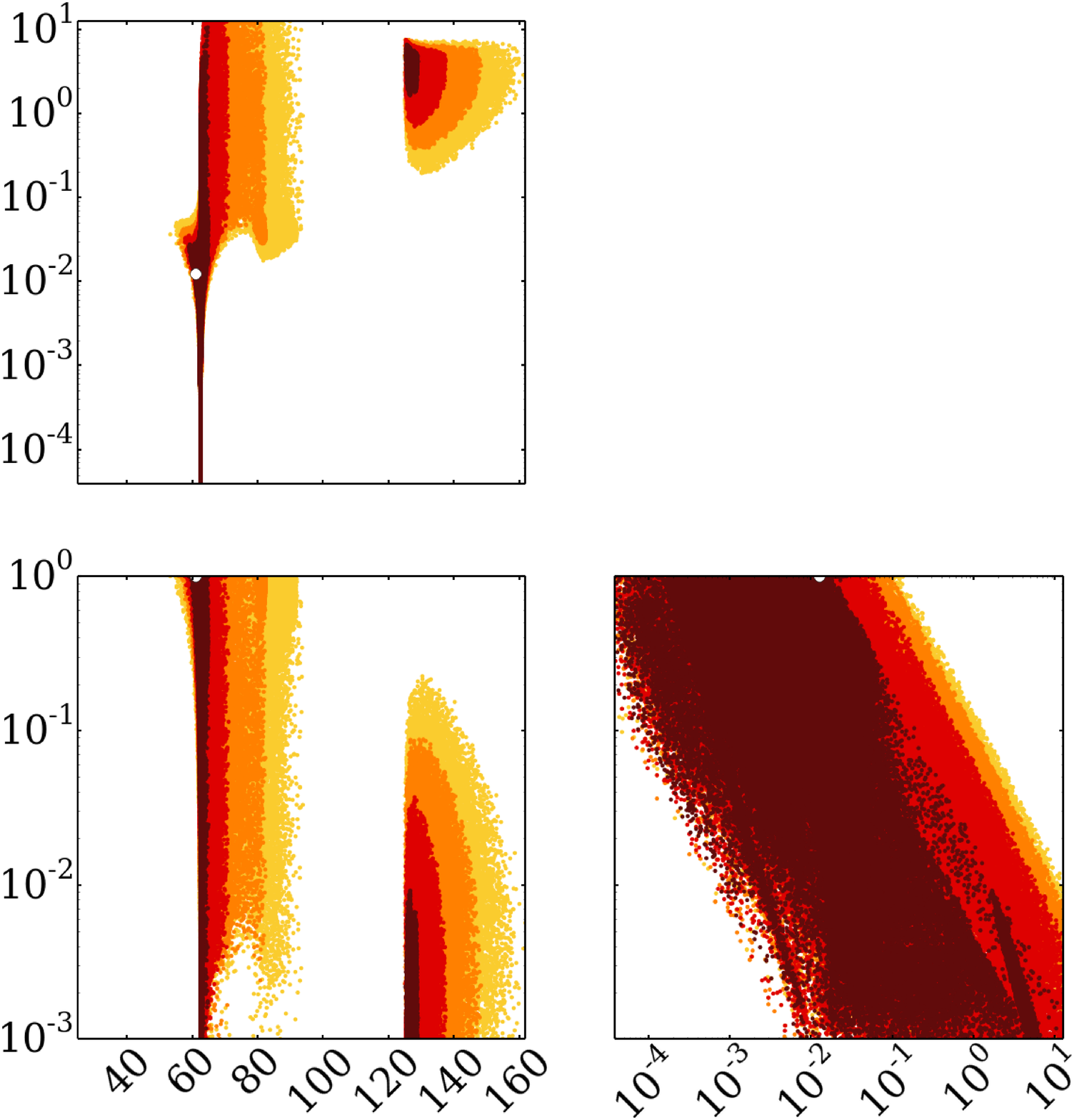}}
  \put(0.225,0.555){\footnotesize \underline{GCE+$\text{BR}_\text{inv}$}}
  \put(0.127,0.01){\footnotesize $m_S \,[\text{GeV}]$}
  \put(0.39,0.01){\footnotesize  $\lamhs$}
  \put(-0.01,0.17){\rotatebox{90}{\footnotesize $R$}}
  \put(-0.01,0.4){\rotatebox{90}{\footnotesize  $\lamhs$}}
  }
 \put(0.52,0.62){ 
  \put(0.0,0.03){\includegraphics[width=0.55\textwidth]{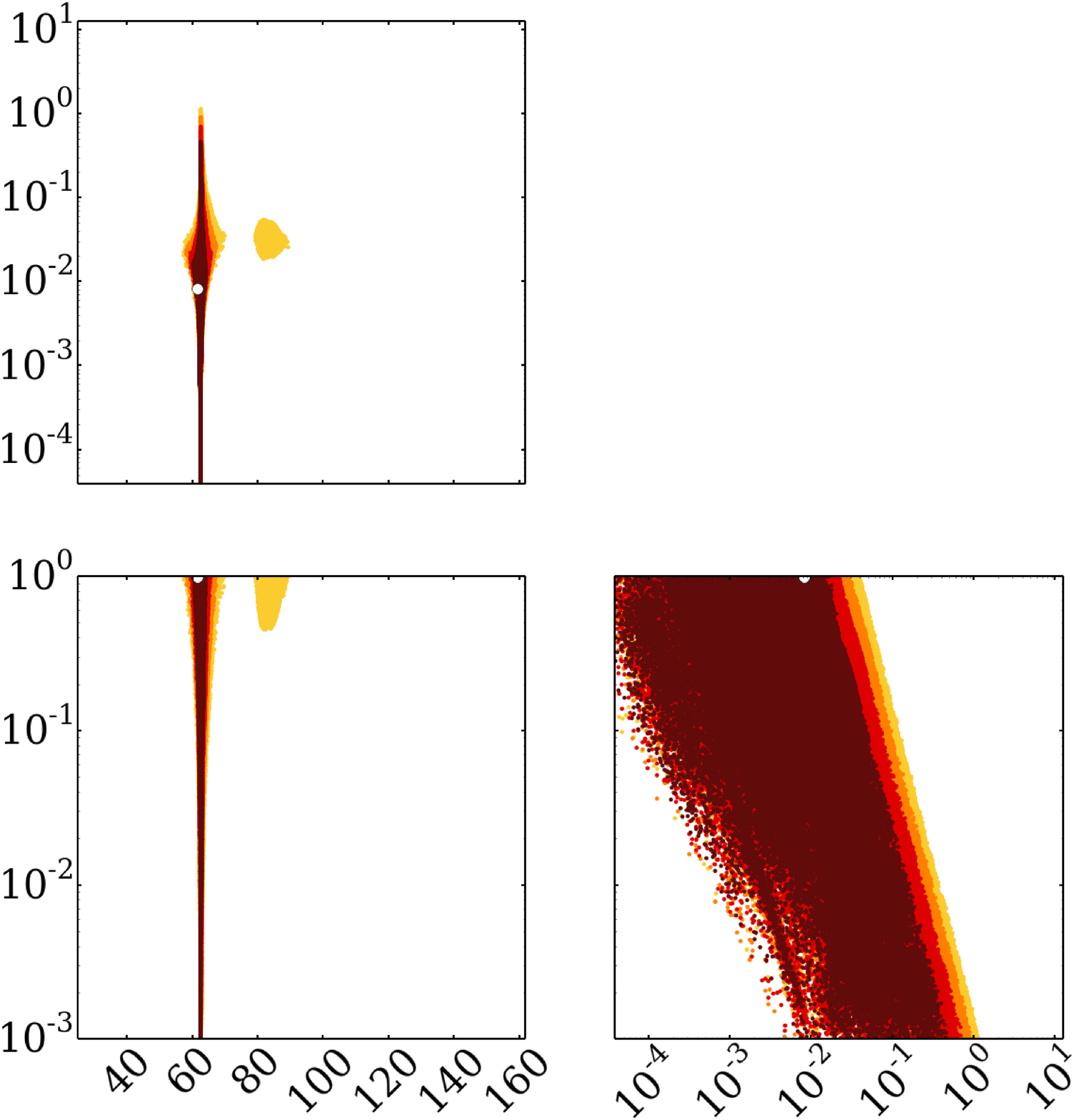}}
  \put(0.191,0.555){\footnotesize \underline{GCE+$\text{BR}_\text{inv}$+LUX}}
  \put(0.127,0.01){\footnotesize $m_S \,[\text{GeV}]$}
  \put(0.39,0.01){\footnotesize  $\lamhs$}
  \put(-0.01,0.17){\rotatebox{90}{\footnotesize $R$}}
  \put(-0.01,0.4){\rotatebox{90}{\footnotesize  $\lamhs$}}
  }
\end{picture}
\vspace*{-5mm}
\caption{Results of a fit to the GCE with free parameters $m_S$, $\lambda_{H\!S}$, $R$ and 
$\logJ$, 
taking into account the log-likelihood from the GCE+$\text{BR}_\text{inv}$ (top left), 
GCE+$\text{BR}_\text{inv}$+LUX (top right), GCE+$\text{BR}_\text{inv}$+LUX+dwarfs 
(bottom left) and GCE+$\text{BR}_\text{inv}$+LUX+dwarfs+lines (bottom right). 
The white dot denotes the best-fit point. The dark-red, red, orange and yellow points lie within the
1, 2, 3 and $4\sigma$ region around the best-fit point, respectively. 
}
\label{fig:Rneq1interm}
\end{figure}

\begin{figure}[h!]
\centering
\setlength{\unitlength}{1\textwidth}
\begin{picture}(1,1.01)
 \put(0.0,0.0){ 
  \put(-0.022,-0.02){\includegraphics[width=1.1\textwidth]{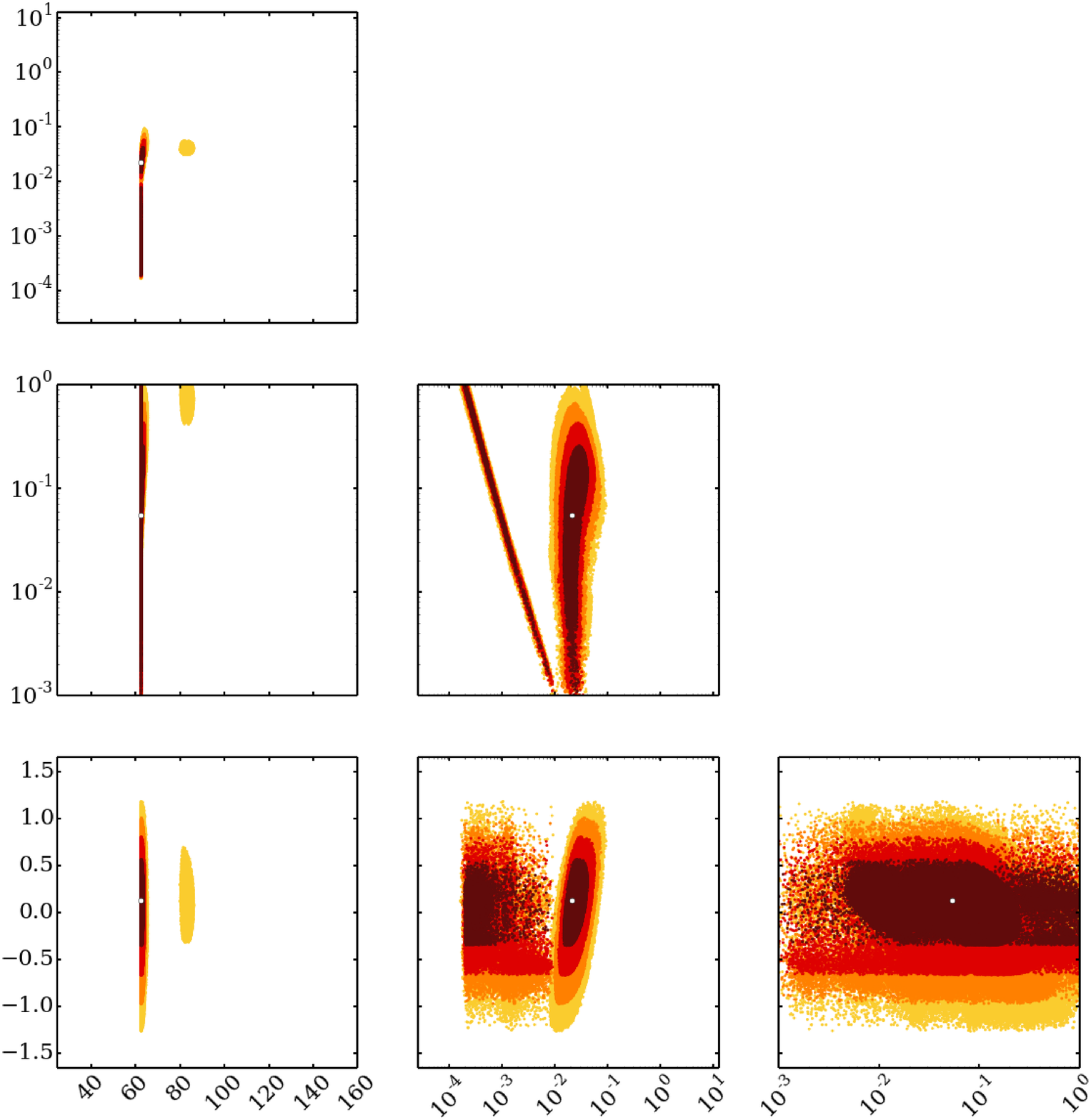}}
  \put(0.204,0.01){\footnotesize $m_S \,[\text{GeV}]$}
  \put(0.527,0.01){\footnotesize $\lamhs$}
  \put(0.838,0.01){\footnotesize $R$}
  \put(0.019,0.135){\rotatebox{90}{\footnotesize $\logJ$}}
  \put(0.019,0.52){\rotatebox{90}{\footnotesize $R$}}
  \put(0.019,0.824){\rotatebox{90}{\footnotesize $\lamhs$}}
  }
\end{picture}
\caption{Results of a fit to the GCE with free parameters $m_S$, $\lambda_{H\!S}$, $R$ and $\logJ$.
The white dot denotes the best-fit point. The dark-red, red, orange and yellow points lie within the
1, 2, 3 and $4\sigma$ region around the best-fit point, respectively. We take into account all constraints
(including the relic density constraint).
}
\label{fig:Rneq1all}
\end{figure}

We now consider the effect of the various constraints discussed in section~\ref{sec:constraints} by including the corresponding likelihoods in the GCE fit. The results of these global fits are presented in figures~\ref{fig:Rneq1interm} and \ref{fig:Rneq1all}; figure~\ref{fig:Rneq1interm} provides information on the viable parameter space in $m_S$, $\lambda_{H\!S}$, $R= \rho_\text{WIMP}/\rho_{\text{total}}$ and $\logJ$,
when adding successively the constraints from the invisible Higgs branching ratio, the LUX direct detection limit, the limits from dwarf satellites and the limits from spectral $\gamma$ lines to the GCE fit. Figure~\ref{fig:Rneq1all} shows detailed information on the results of a global fit to the GCE signal with all constraints added, including in particular the DM relic density.  We shall now discuss the impact of the various constraints in turn.

\begin{itemize}

\item DM masses below the Higgs resonance, $m_S < m_h/2$, lead to invisible Higgs decays, $h\to SS$, and are thus constrained by the LHC limit on the invisible Higgs branching ratio, see section~\ref{sec:BRinv}. The limit on invisible Higgs decays  cuts out the region $\lambda_{H\!S} \gtrsim 0.02$ for $m_S \lesssim m_h/2$,  and the best fit point for the GCE signal moves to the resonant region, $m_S \approx m_h/2$, see figure~\ref{fig:Rneq1interm} upper left panel. The parameter region above the Higgs threshold, $m_S \gtrsim m_h$, is still viable and lies within $1\sigma$ of the best fit point.

\item Severe constraints on the model parameter space are imposed by the current direct detection limits from LUX~\cite{Akerib:2013tjd}, see section~\ref{sec:LUX}. These limits exclude the parameter space corresponding to 
$\lambda_{H\!S} \gtrsim 0.02$ (for $R=1$) and $\lambda_{H\!S} \gtrsim 0.5$ (for $R\simeq10^{-3}$),
and in particular remove the Higgs threshold region. 
 As shown in the upper right panel of figure~\ref{fig:Rneq1interm}, only the
region near the resonance, $m_S \approx m_h/2$, and a small and barely viable parameter region at the $W$-threshold survives. The region at the $W$-threshold is, however, already between 3 and $4\sigma$ away from the best-fit point.

\item  Adding the constraints from dwarf galaxy  $\gamma$-ray searches as described in section~\ref{sec:constraints_indirect} does not significantly modify the viable range of the model parameters $m_S$ and $\lambda_{H\!S}$, see figure~\ref{fig:Rneq1interm} lower left panel. However, a light tension between the dwarf limits and the flux required for the GCE signal is present, and  the fit thus prefers a larger 
$\logJ$. A larger $ J_{40^\circ}$ allows for a smaller $\sigmav$, whilst still maintaining the correct GCE flux. Adding also searches for spectral $\gamma$ lines reduces the viable $\logJ$ further, as shown in the lower right panel of figure~\ref{fig:Rneq1interm}. Values $\logJ \gtrsim 1$ are now disfavoured, as a larger $ J_{40^\circ}$ implies an even larger value for the $J$-factors of the $\gamma$-lines, $ J_{3^{\circ}}$ and $ J_{16^\circ}$, see figure~\ref{fig:Jbar2}.  The fit presented in figure~\ref{fig:Rneq1interm} takes into account the log-likelihood from R3 which 
is optimised for a `cuspy' dark matter profile as considered here. Indeed we 
found that R3 provides a stronger constraint than R16 in the considered region of parameter space.

\end{itemize}

Finally, we discuss the constraint from the DM relic density, see section~\ref{sec:reldens}. 
Requiring $\Omega h^2|_\text{DM,\,total}= \Omega h^2|_\text{WIMP}/R = \Omega h^2|_\text{Planck}$
further constrains the parameter space and leads to interesting parameter correlations.
Note, however, that the connection between the DM annihilation cross section $\sigmav$ and the relic density $\Omega h^2$ is based on 
the assumption of a standard cosmological history. Deviations from the standard cosmological scenario could lead to both a reduction or an enhancement of the predicted relic density. 
In order to understand this structure we first consider the more simple case $R=1$, and then the general case $R\leq1$.

\begin{itemize}

\item If we require  that the scalar Higgs portal WIMP constitutes all of DM, i.e., $R = 1$,
only a small viable region remains with $\lambda_{H\!S} \approx 2 \times 10^{-4}$ near the very tip of the resonance at $m_S \approx m_h/2$. 
At this point of the parameter space, the annihilation cross sections as of today, $\sigmav_\text{today}$, and at the time of freeze out, $\sigmav_\text{freeze-out}$, are of the same order, and the flux required to describe the GCE and the cosmological dark matter relic density 
can be both accommodated at the same time.
For dark matter masses slightly above the resonance tip, however, the ratio $\sigmav_\text{today}/\sigmav_\text{freeze-out}$ increases rapidly, 
since the annihilation cross section $\sigmav$ exhibits a strong velocity dependence in the vicinity of the $s$-channel Higgs resonance. 
Thus, for $m_S \gtrsim m_h/2$,  the flux required to describe the GCE implies a reduced $\sigmav_\text{freeze-out}$ and in turn dark matter relic densities which exceed the observed cosmological abundance, so that $R=1$  is not possible.   
Further away from the resonance  $\sigmav_\text{today}/\sigmav_\text{freeze-out}$ approaches values of order one again, 
and a second region of parameter space opens up where it is possible to reconcile the GCE flux and the DM relic density. 
This region, however, is in tension with direct detection constraints. 

\item The possibility of a non-WIMP dark matter component, corresponding to $R < 1$, opens up more parameter space. The increase in the ratio $\sigmav_\text{today}/\sigmav_\text{freeze-out}$, when moving from the very tip of the resonance at $m_S \approx m_h/2$ to slightly larger dark matter masses, can be compensated by a decrease in $R$. 
Since  $\Omega_\text{DM,\,total}  \propto \Omega_\text{WIMP}/R \propto 1/(R\, \sigmav_\text{freeze-out})$,
and $\sigmav_\text{freeze-out} \propto  \lambda_{H\!S}^2$,
requiring $\Omega_\text{Planck} = \Omega_\text{DM,\,total}$ 
implies that $R\propto 1/\lambda_{H\!S}^2$ for a given mass. This anti-correlation between $R$ and $\lamhs^2$ is clearly visible as the narrow stripe 
in the $R$-$\lambda_{H\!S}$ panel of figure~\ref{fig:Rneq1all}.

As mentioned above, slightly further away from the resonance, $\sigmav_\text{today}/\sigmav_\text{freeze-out}$ approaches values of order one again, and thus $R\approx 1$ would be viable. 
However, due to the direct detection constraint which disfavours large $R \times \lamhs$, the region with $R = 1$ is already $3\sigma$ away from the best fit point and $R\lesssim 0.5$ is preferred in this region. It is interesting to note that the viable regions of parameter space extend down to $R\approx 10^{-3}$, i.e., the GCE signal can be explained by a thermal relic WIMP that only constitutes one per mille of the dark matter.

\end{itemize}

\begin{figure}[h!]
\centering
\setlength{\unitlength}{1\textwidth}
\begin{picture}(0.99,0.54)
 \put(0.0,0.0){ 
    \put(-0.02,0.03){\includegraphics[width=1.1\textwidth]{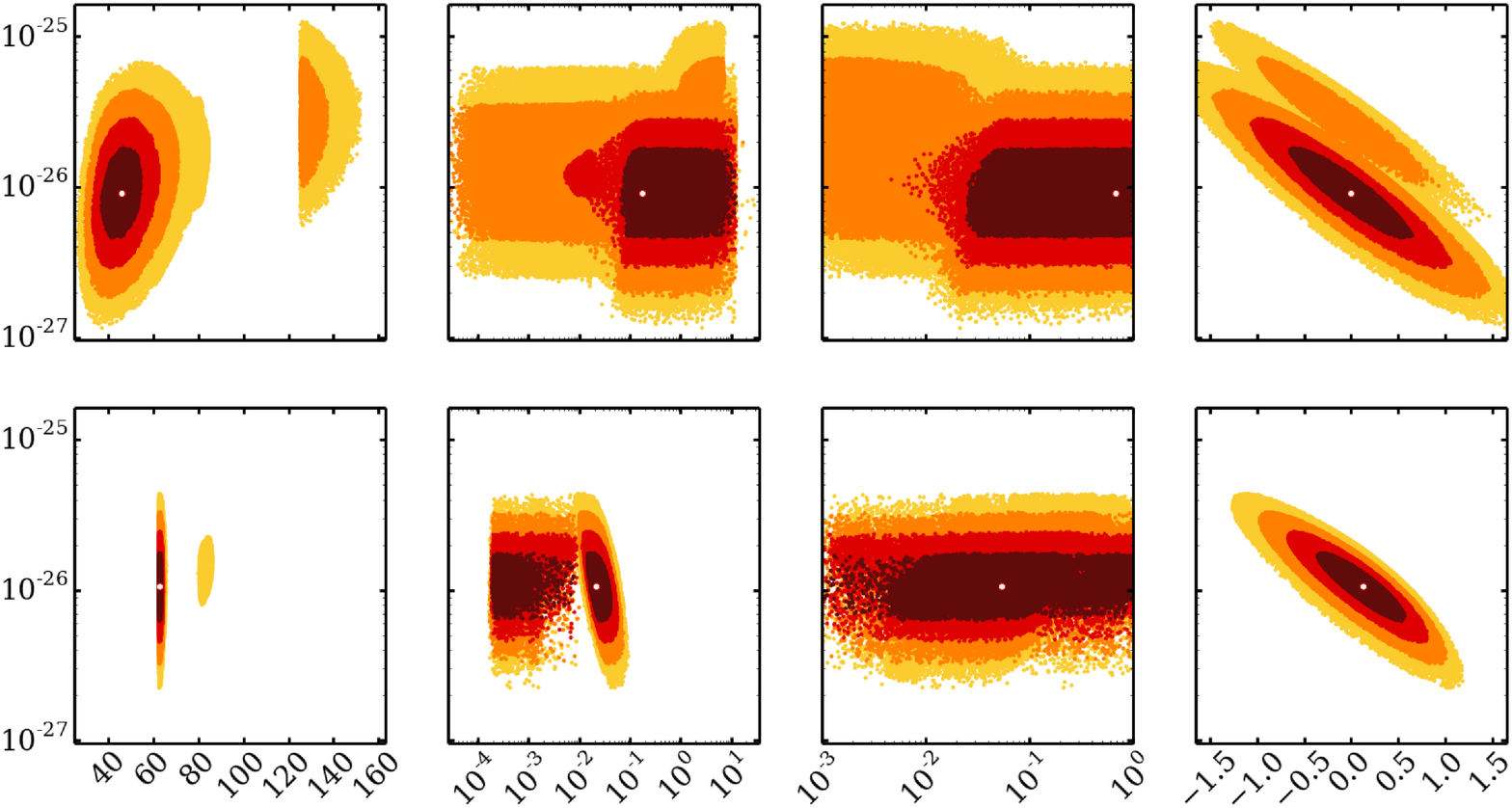}}
  \put(0.17,0.01){\footnotesize $m_S \,[\text{GeV}]$}
  \put(0.42,0.01){\footnotesize $\lamhs$}
  \put(0.643,0.01){\footnotesize $R$}
  \put(0.801,0.01){\footnotesize $\logJ$}
  \put(0.025,0.108){\rotatebox{90}{\footnotesize $\sigmav\, R^2 \,[\text{cm}^3\,\text{s}^{-1}]$}}
  \put(0.025,0.345){\rotatebox{90}{\footnotesize $\sigmav\, R^2 \,[\text{cm}^3\,\text{s}^{-1}]$}}
  }
\end{picture}
\caption{The correlation between the annihilation cross section as of today, $\sigmav$, and  the input parameters of the fit, $m_S$, $\lambda_{H\!S}$, $R$ and $\logJ$ for the case of the GCE-only fit (upper row) and taking into account all the constraint, including the relic density constraint  (lower row).
The white dot denotes the best-fit point. The dark-red, red, orange and yellow points lie within the
1, 2, 3 and $4\sigma$ region around the best-fit point, respectively.}
\label{fig:Rneq1sigmav}
\end{figure}

To describe the GCE $\gamma$-ray flux from the annihilation of the scalar WIMP, the cross section  has to be $\sigmav \times R^2 \approx 1 \times 10^{-26}{\rm cm}^3/{\rm s}$ . The correlation between $\sigmav \times R^2$ and the four fit parameters 
$m_S$, $\lambda_{H\!S}$, $R$ and $\logJ$ is shown in figure~\ref{fig:Rneq1sigmav}  for the case of the GCE-only fit (upper panels) and the fit
including all the constraints (lower panels). Adding the constraints reduces the viable region of model parameter space to scalar masses $m_S \approx m_H/2$ and the possible values of Higgs-scalar coupling to $\lambda_{H\!S} \lesssim 0.1$. Furthermore, the allowed range in $\logJ$ is partly reduced.
As expected there is a strong anti-correlation between $\sigmav \times R^2$ and $\logJ$.
A smaller $\sigmav \times R^2$ is allowed for a larger  $\logJ$, and vice-versa.

\begin{figure}[h!]
\centering
\setlength{\unitlength}{1\textwidth}
\begin{picture}(0.7,0.4)
 \put(0.0,0.0){ 
  \put(0.0,-0.005){\includegraphics[width=0.72\textwidth]{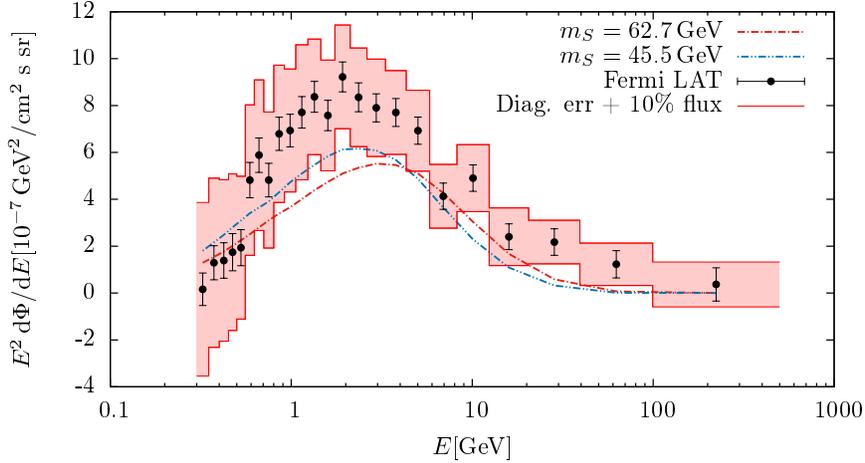}}
  }
\end{picture}
\caption{Spectral fit for the best fit point taking in account all constrains (red) and fitting only the GCE (blue). The red band displays the diagonal part of the covariance matrix including a 10\% error on the predicted flux (red curve). The black points are the {\it Fermi}-LAT observation including an uncorrelated statistical error.
}
\label{fig:bestfitspec}
\end{figure}

Besides the overall flux, the model also has to accommodate the spectral shape of the GCE. In figure~\ref{fig:bestfitspec} we show the energy spectrum of the $\gamma$-ray flux for the GCE-only fit and the fit
including all the constraints, compared to the {\it Fermi}-LAT  data as analysed in \cite{Calore:2014xka}. The correlated systematic errors from astrophysical uncertainties are shown as the red shaded areas, while the error bars denote the uncorrelated statistical errors, see \cite{Calore:2014xka} for details. Both fits predict $\gamma$-ray spectra which are systematically lower than the mean values of the GCE spectrum. However, as the systematic astrophysical uncertainties (red shaded areas) are strongly correlated, both the GCE only fit and the fit including all constraints provide a good description of the GCE energy spectrum, as discussed in more detail below. Note that the systematic errors depend on the theory prediction as explained in section~\ref{sec:gcefit}. In Fig.~\ref{fig:bestfitspec} 
we show the errors for the fit including all constraints although the difference to the GCE-only fit would be barely noticeable in the figure.

In table~\ref{tab:res} we have collected the best fit values for the scalar Higgs portal model parameters $m_S$, $\lambda_{H\!S}$, and for the astrophysical parameters $R= \rho_\text{WIMP}/\rho_{\text{total}}$ and $ J_{40^\circ}$.   We show results for the various global fits, including the GCE signal only, and for the GCE signal with the different constraints added successively. As discussed above, the scalar Higgs portal model can describe the GCE signal for dark matter masses near $m_S = m_h/2$, and for perturbative values of the Higgs-scalar coupling, $\lambda_{H\!S}\lesssim10^{-2}$. After taking into account all constraints two viable regions of parameter space emerge: one region where $\lambda_{H\!S}\approx 2 \times 10^{-2}$ and where an additional dark matter component beyond the scalar WIMP of our model is preferred ($R\lesssim 0.5$), and one region 
where the scalar WIMP constitutes all of dark matter, $R \lesssim 1$, and where $\lambda \approx 
 2\times 10^{-4} \lesssim \lambda_{H\!S} \lesssim 10^{-2}$. For both these regions, the best fit point has $\chi^2 \approx 27$ with respect to the GCE signal. For the  25 data points and the 4 parameters, this corresponds to a $p$-value of $p=0.18$, and thus indicates a reasonably good fit to the GCE. 
 
 \begin{sidewaystable}[p]
\begin{center}
\renewcommand{\arraystretch}{1.5}
\begin{tabular}{c | c  c  c  c  c  c  c} 
$\log L$ contribution  & GCE  & +$\text{BR}_{\text{inv}}$  & +LUX  & +dwarfs  & +lines  & +relic den.  & 2nd region\\ \hline\hline 
$m_{S}$\,{[}GeV{]}  & $45.50_{-5.36}^{+5.98}$  & $61.07_{-1.98}^{+2.65}$ & $61.55_{-0.85}^{+1.78}$ & $61.35_{-0.79}^{+1.90}$ & $61.46_{-0.85}^{+1.87}$  & $62.70_{-0.18}^{+0.57}$ & $62.52_{-0.01}^{+0.02}$\\
$\lamhs$ & $0.17_{-0.09}^{+11.67}$  & $0.0125_{-0.0125}^{+7.31}$  & $0.0082_{-0.0082}^{+0.317}$  & $0.0087_{-0.0087}^{+0.312}$  & $0.0082_{-0.0082}^{+0.315}$  & $0.022_{-0.013}^{+0.015}$ & $0.00029_{-0.00010}^{+0.0078}$\\
$R$ & $0.68_{-0.65}^{+0.32}$  & $1.0_{-1.0}^{+0.0}$  & $0.99_{-0.99}^{+0.01}$  & $1.0_{-1.0}^{+0.0}$  & $1.0_{-1.0}^{+0.0}$  & $0.054_{-0.053}^{+0.141}$ & $0.498_{-0.496}^{+0.502}$\\
$\log J/J_{\text{nom}}$ & $0.0_{-0.44}^{+0.44}$  & $-0.05_{-0.36}^{+0.48}$  & $0.02_{-0.43}^{+0.42}$  & $0.22_{-0.35}^{+0.36}$  & $0.12_{-0.29}^{+0.31}$  & $0.13_{-0.32}^{+0.30}$ & $0.13_{-0.31}^{+0.32}$\\
$\sigma v\,[10^{-26}\,\text{cm}^{3}/\text{s}]$  & $1.97_{-1.38}^{+1034}$  & $1.28_{-0.61}^{+4.1\textrm{e}6}$ & $1.23_{-0.55}^{+1.7\textrm{e}6}$  & $0.96_{-0.37}^{+1.3\textrm{e}6}$  & $1.04_{-0.42}^{+1.3\textrm{e}6}$  & $359_{-327}^{+9.7\textrm{e}5}$ & $4.3_{-0.9}^{+1.6\textrm{e}5}$\\
$\sigma v\, R^{2}\,[10^{-26}\,\text{cm}^{3}/\text{s}]$  & $0.91_{-0.35}^{+0.53}$  & $1.28_{-0.53}^{+2.02}$ & $1.21_{-0.45}^{+0.68}$  & $0.96_{-0.31}^{+0.43}$  & $1.04_{-0.32}^{+0.39}$  & $1.06_{-0.32}^{+0.42}$ & $1.06_{-0.31}^{+0.43}$\\
$\chi_{\text{GCE}}^{2}$  & 19.3  &  25.3  & 25.6  & 26.0 & 26.0  & 26.8 & 26.7\\
$p(\chi_{\text{GCE}}^{2})$  & 0.57  &  0.20 & 0.24  & 0.22  & 0.21  & 0.18 & 0.18\\
$p(\text{BR}_{\text{inv}})$ & 0.0  & 0.90 & 0.97 & 0.97 & 0.97 & 1.0 & 1.0\\
$p$(LUX) & 0.0  & 0.32  & 0.62  & 0.58  & 0.62  & 0.84 & 1.0\\
$p$(dwarfs)  & 0.18  & 0.16 & 0.18  & 0.24  & 0.22 & 0.22 & 0.22\\
$p$(lines R3)  & 0.5  & 0.5  & 0.5  & 0.5 & 0.5  & 0.5  & 0.5\\
$p$(relic den.)  & 0.03  & 0.0 & 0.0  & 0.0  & 0.0  & 0.99 &1.0\\
\end{tabular}
\renewcommand{\arraystretch}{1}
\end{center}
\caption{Best fit points and corresponding 1$\sigma$ error for fits to the GCE only, and including successively constraints from the invisible Higgs branching ratio, direct detection limits, independent searches for $\gamma$-rays from dwarf satellite galaxies, searches for spectral $\gamma$ lines, and from the dark matter relic density. We also display the best fit in a second, viable region of parameter space (last column). 
Also shown are the $\chi^2_\text{GCE}$ and the $p$-values of the respective best-fit points taking into account
the log-likelihood contributions of the observables given in the first line. 
$p(\chi^2_\text{GCE})$ represents the goodness of the GCE fit for 25 data points and 4 fitted parameters.
The remaining $p$-values represent the confidence level at which the best fit
is compatible with the constraints coming from each extra-observable we include in the fit (see text for more details). 
}
\label{tab:res}
\end{sidewaystable}

In table~\ref{tab:res} we also list the $p$-values corresponding to all the constraints we use, i.e., more precisely, the confidence level at which the best fit
is compatible with the constraints coming from each single additional observable (BR$_{\rm inv}$, LUX, dwarfs, lines and DM relic density) we include in the fit.%
\footnote{The $p$-values are calculated considering  the likelihood of the single observable $\mathcal{L}_{i}$,
and finding the contour with respect to the maximum $\mathcal{L}_{i}$ 
which passes through the point corresponding to the best-fit when including all the constraints.
The $p$-value is the confidence level of the contour.
To this purpose, since all the single observables do not depend
on $J_{40^\circ}$ and depend on $R$ only as a rescaling parameter, we assume that $\mathcal{L}_{i}$
 follows a $\chi^2$ distribution with two degrees of freedom,
i.e., that it depends only on the two parameters $m_S$ and a combination of $R$ and $\lambda_{H\!S}$.
The  $\gamma$-ray lines signal, instead, depends on $J_{40^\circ}$, but, again, only as a rescaling parameter.
We thus assume a $\chi^2$ distribution with two degrees of freedom also for the lines $\mathcal{L}_{i}$.}
We find that for both viable regions of parameter space the GCE global fit is well compatible with all the constraints. A light tension is present only with respect to the limits from the dwarf galaxies; 
the corresponding $p$-value of $p=0.23$, is however well within an acceptable range.
Note, nonetheless, that we have assumed nominal values from \cite{Ackermann:2015zua,Ahnen:2016qkx} for the uncertainties in the $J$-factors of the dwarfs.
On the other hand,  it has been pointed out that this uncertainty has been possibly underestimated \cite{Bonnivard:2015xpq}.
This would contribute to relax the dwarf constraints and alleviate the mild tension.

\begin{figure}[h!]
\centering
\setlength{\unitlength}{1\textwidth}
\begin{picture}(1,0.35)
 \put(0.0,0.0){ 
  \put(0.02,0.01){\includegraphics[width=0.31\textwidth]{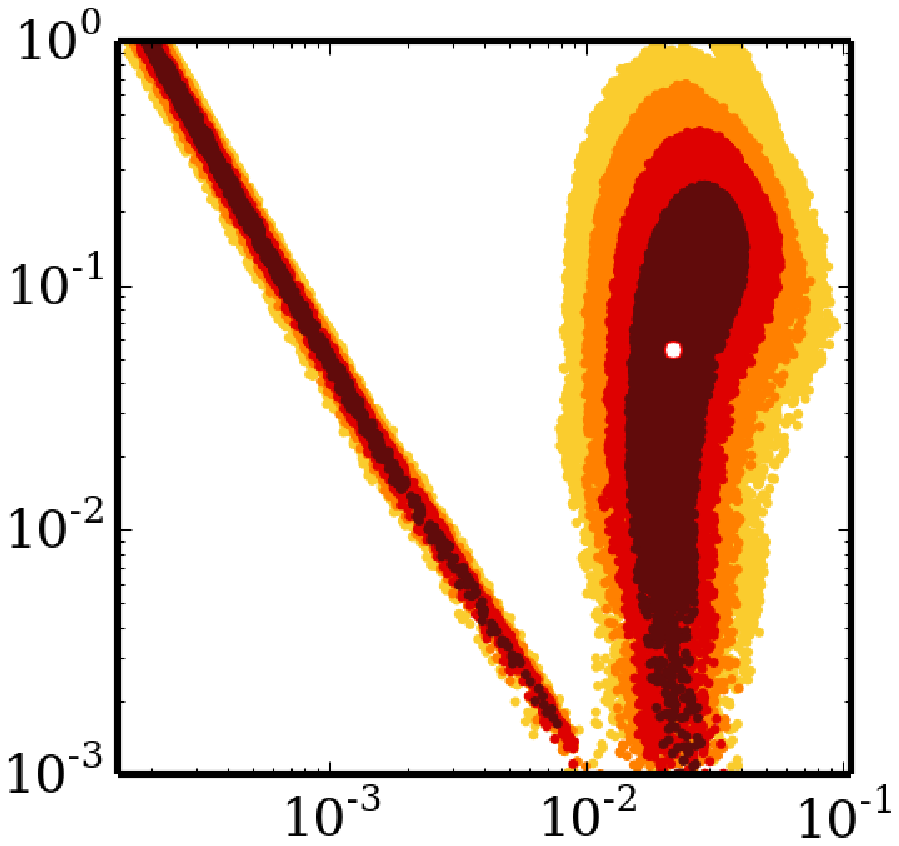}}
  \put(0.17,0.01){\footnotesize $\lamhs$}
  \put(0.0,0.17){\rotatebox{90}{\footnotesize $R$}}
}
 \put(0.34,0.0){ 
  \put(0.02,0.01){\includegraphics[width=0.31\textwidth]{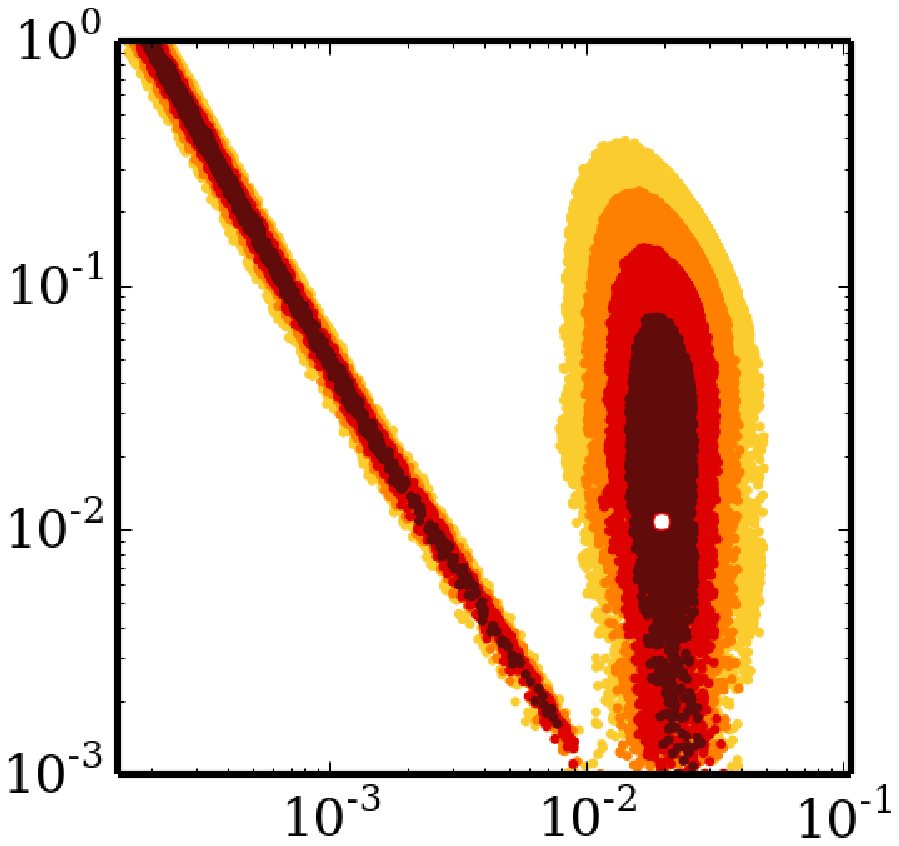}}
  \put(0.17,0.01){\footnotesize $\lamhs$}
  \put(0.0,0.17){\rotatebox{90}{\footnotesize $R$}}
}
 \put(0.68,0.0){ 
  \put(0.02,0.01){\includegraphics[width=0.31\textwidth]{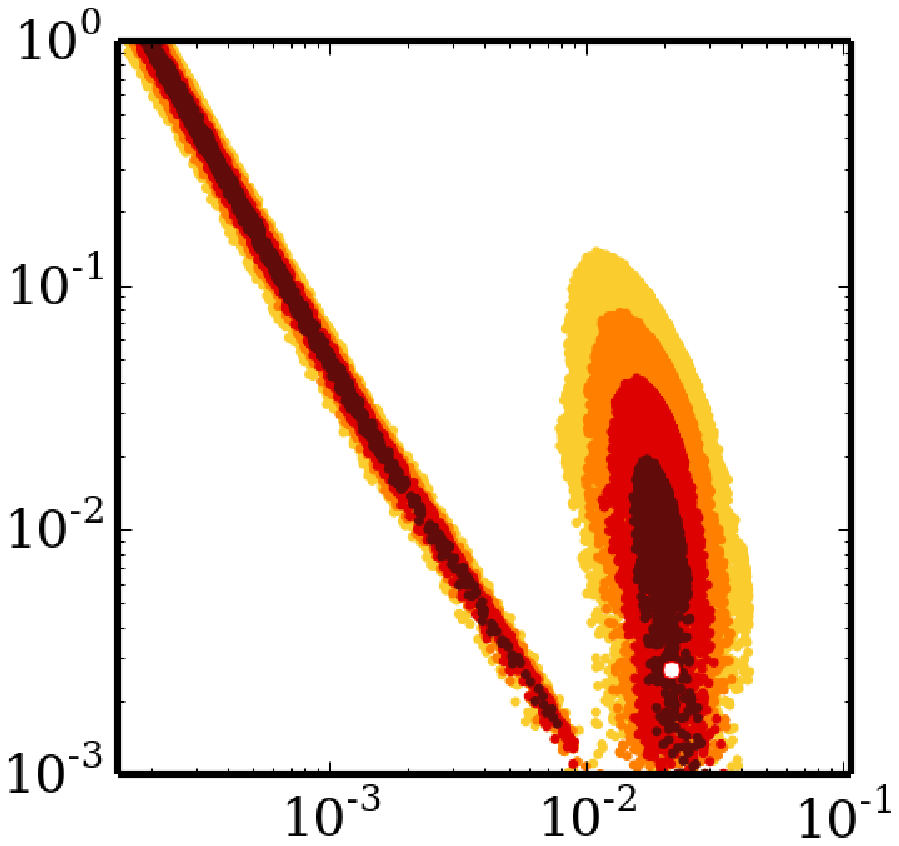}}
  \put(0.17,0.01){\footnotesize $\lamhs$}
  \put(0.0,0.17){\rotatebox{90}{\footnotesize $R$}}
  }
\end{picture}
\caption{Results of a fit to the GCE with free parameters $m_S$, $\lambda_{H\!S}$, $R$ and $\logJ$. We take into account 
all constraints (including the relic density constraint) and focus on the correlation between $R$ and $\lamhs$. 
In the middle and right panel we include potential future direct detection limits with a sensitivity of 10 and 50 times the current LUX sensitivity, respectively. The white dot denotes the best-fit point. The dark-red, red, orange and yellow points lie within the
1, 2, 3 and $4\sigma$ region around the best-fit point, respectively.
}
\label{fig:Rneq1DDproj}
\end{figure}

Finally, we have studied the impact of future direct detection experiments like XENON1T~\cite{Aprile:2015uzo} or DARWIN~\cite{Baudis:2012bc}. We simply assume a sensitivity to $\sigmav$ which is 10 or 50 times larger than that of the current LUX limits, and include those potential future limits in our global fit. The result is displayed in figure~\ref{fig:Rneq1DDproj} where we show the viable regions in the $R$-$\lambda_{H\!S}$ plane given the current LUX bounds (left panel), and a 10 times (middle panel) and 50 times (right panel) larger potential future sensitivity. The future direct detection experiments will probe part of the currently allowed region of parameter space, and stronger limits would exclude regions with both $\lambda_{H\!S}$ and $R$ large. On the other hand, even limits corresponding to a 50 times larger sensitivity than that of current direct detection experiments would leave viable regions of model parameter space, corresponding to either sizeable  Higgs-scalar couplings $\lambda_{H\!S}\approx 2 \times 10^{-2}$ and small $R\lesssim 0.1$, or to smaller $\lambda_{H\!S}$ but values of $R$ close to one. Note that future limits from CTA~\cite{Consortium:2010bc} will be relevant for DM masses above about 100~GeV, see e.g.~\cite{Silverwood:2014yza,Cline:2013gha,Beniwal:2015sdl}, and would thus not constrain the GCE interpretation within the scalar Higgs portal model further. Also LHC searches for scalar Higgs portal models with dark matter masses $m_S \gtrsim m_h/2$ are not sensitive to the relevant model parameter space, see \cite{Djouadi:2012zc,Endo:2014cca,Craig:2014lda,Han:2016gyy}. 
The best test of these two regions will probably come from further $\gamma$-ray searches. 
Limits from both dwarf galaxies and lines are expected to improve with time while larger statistics is being collected. In addition, limits from dwarfs
will further improve since more dwarf galaxies, potentially with large $J$-factors, should be discovered in the next years by 
current surveys like DES~\cite{Abbott:2005bi}, and future ones like LSST~\cite{Ivezic:2008fe}.  This will clarify if the present tension with the GCE will become
more severe at the point of excluding the remaining parameter space, or, if, eventually, a signal in dwarfs and lines will 
emerge, confirming the GCE interpretation.

\section{Conclusion}\label{sec:summary} 

We have presented a global fit of the $\gamma$-ray galactic center excess (GCE) within a minimal Higgs portal model with a scalar WIMP dark matter particle. We find that the 
shape and strength of the GCE is well described by dark matter annihilation in various regions of parameter space, including in particular the resonance and threshold 
regions where the dark matter mass, $m_S$, is close to (half the) Higgs and $W$ boson masses, respectively.

The parameter space of the scalar Higgs portal model is constrained by the search for invisible Higgs decays, direct dark matter searches, searches for dark matter annihilation from dwarf spheroidal galaxies, 
searches for mono-energetic spectral $\gamma$-lines from the Milky Way halo, and by the cosmological dark matter relic abundance. 
We have included these constraints into the global fit of the GCE signal and studied the implications for the model parameter space, taking properly into account the 
theoretical uncertainty from the dark matter distribution. Furthermore, we consider the possibility that the dark sector is more complex than that of the minimal Higgs portal model  
and allow for scalar WIMP dark matter densities smaller than the total, gravitationally interacting dark matter density. With this freedom, we can easily accommodate the GCE signal in 
regions of model parameter space which would otherwise be excluded because the model prediction would exceed the cosmologically observed dark mater relic density. 

Taking into account all constraints, the scalar Higgs portal model can describe the GCE signal if the dark matter mass is near $m_S = m_h/2$, so that dark matter annihilation proceeds through resonant Higgs exchange. 
Two regions of parameter space are viable: one region where the Higgs--dark-matter coupling, $\lambda_{H\!S}$, is of order ${\cal O}(10^{-2})$ and where an additional dark matter component beyond a scalar WIMP is preferred, and one region where $\lambda_{H\!S}$ may be significantly smaller, but where the scalar WIMP constitutes a significant fraction or even all of dark matter. 
These regions emerge from an interplay between the different scaling of the GCE signal and the relic density with the fraction of WIMP dark matter, $R= \rho_\text{WIMP}/\rho_{\text{total}}$, and the 
strong velocity dependence of the annihilation cross section near the resonance. Note that this effect can potentially be of relevance for every model with resonant annihilation.
The favoured regions of scalar dark matter masses and couplings are hard to probe in future direct detection and collider experiments. Future searches for $\gamma$-ray emission from dwarf spheroidal galaxies, however, will be able to confirm or exclude the dark matter interpretation of the galactic center excess.

\section*{Acknowledgements}

We would like to thank Andrea Albert, Thejs Brinckmann, Marco Cirelli, Michael Korsmeier 
and Julien Lesgourgues for helpful discussions. 
We acknowledge support by the German Research Foundation DFG through the research unit ``New physics at the LHC'', the 
Helmholtz Alliance for Astroparticle Physics and the German Federal Ministry of Education 
and Research BMBF. 

\addcontentsline{toc}{section}{References}
\bibliographystyle{utphys.bst}
\bibliography{GCE_fits_ref}

\providecommand{\href}[2]{#2}\begingroup\raggedright\begin{thebibliography}{10}

\bibitem{Bertone:2004pz}
G.~Bertone, D.~Hooper, and J.~Silk, {\em {Particle dark matter: Evidence,
  candidates and constraints}}.
  \href{http://dx.doi.org/10.1016/j.physrep.2004.08.031}{Phys.Rept. {\bf 405}
  (2005)  279--390},
\href{http://arxiv.org/abs/hep-ph/0404175}{{\tt arXiv:hep-ph/0404175
  [hep-ph]}}.

\bibitem{Drees:2012ji}
M.~Drees and G.~Gerbier, {\em {Mini-Review of Dark Matter: 2012}}.
\href{http://arxiv.org/abs/1204.2373}{{\tt arXiv:1204.2373 [hep-ph]}}.

\bibitem{Klasen:2015uma}
M.~Klasen, M.~Pohl, and G.~Sigl, {\em {Indirect and direct search for dark
  matter}}. \href{http://dx.doi.org/10.1016/j.ppnp.2015.07.001}{Prog. Part.
  Nucl. Phys. {\bf 85} (2015)  1--32},
\href{http://arxiv.org/abs/1507.03800}{{\tt arXiv:1507.03800 [hep-ph]}}.

\bibitem{Silveira:1985rk}
V.~Silveira and A.~Zee, {\em {SCALAR PHANTOMS}}.
\href{http://dx.doi.org/10.1016/0370-2693(85)90624-0}{Phys. Lett. {\bf B161}
  (1985)  136}.

\bibitem{McDonald:1993ex}
J.~McDonald, {\em {Gauge singlet scalars as cold dark matter}}.
  \href{http://dx.doi.org/10.1103/PhysRevD.50.3637}{Phys. Rev. {\bf D50} (1994)
   3637--3649},
\href{http://arxiv.org/abs/hep-ph/0702143}{{\tt arXiv:hep-ph/0702143
  [HEP-PH]}}.

\bibitem{Burgess:2000yq}
C.~P. Burgess, M.~Pospelov, and T.~ter Veldhuis, {\em {The Minimal model of
  nonbaryonic dark matter: A Singlet scalar}}.
  \href{http://dx.doi.org/10.1016/S0550-3213(01)00513-2}{Nucl. Phys. {\bf B619}
  (2001)  709--728},
\href{http://arxiv.org/abs/hep-ph/0011335}{{\tt arXiv:hep-ph/0011335
  [hep-ph]}}.

\bibitem{Goodenough:2009gk}
L.~Goodenough and D.~Hooper, {\em {Possible Evidence For Dark Matter
  Annihilation In The Inner Milky Way From The Fermi Gamma Ray Space
  Telescope}}.
\href{http://arxiv.org/abs/0910.2998}{{\tt arXiv:0910.2998 [hep-ph]}}.

\bibitem{Hooper:2010mq}
D.~Hooper and L.~Goodenough, {\em {Dark Matter Annihilation in The Galactic
  Center As Seen by the Fermi Gamma Ray Space Telescope}}.
  \href{http://dx.doi.org/10.1016/j.physletb.2011.02.029}{Phys. Lett. {\bf
  B697} (2011)  412--428},
\href{http://arxiv.org/abs/1010.2752}{{\tt arXiv:1010.2752 [hep-ph]}}.

\bibitem{Hooper:2011ti}
D.~Hooper and T.~Linden, {\em {On The Origin Of The Gamma Rays From The
  Galactic Center}}. \href{http://dx.doi.org/10.1103/PhysRevD.84.123005}{Phys.
  Rev. {\bf D84} (2011)  123005},
\href{http://arxiv.org/abs/1110.0006}{{\tt arXiv:1110.0006 [astro-ph.HE]}}.

\bibitem{Abazajian:2012pn}
K.~N. Abazajian and M.~Kaplinghat, {\em {Detection of a Gamma-Ray Source in the
  Galactic Center Consistent with Extended Emission from Dark Matter
  Annihilation and Concentrated Astrophysical Emission}}.
  \href{http://dx.doi.org/10.1103/PhysRevD.86.083511,
  10.1103/PhysRevD.87.129902}{Phys. Rev. {\bf D86} (2012)  083511},
  \href{http://arxiv.org/abs/1207.6047}{{\tt arXiv:1207.6047 [astro-ph.HE]}}.
[Erratum: Phys. Rev.D87,129902(2013)].

\bibitem{Hooper:2013rwa}
D.~Hooper and T.~R. Slatyer, {\em {Two Emission Mechanisms in the Fermi
  Bubbles: A Possible Signal of Annihilating Dark Matter}}.
  \href{http://dx.doi.org/10.1016/j.dark.2013.06.003}{Phys. Dark Univ. {\bf 2}
  (2013)  118--138},
\href{http://arxiv.org/abs/1302.6589}{{\tt arXiv:1302.6589 [astro-ph.HE]}}.

\bibitem{Gordon:2013vta}
C.~Gordon and O.~Macias, {\em {Dark Matter and Pulsar Model Constraints from
  Galactic Center Fermi-LAT Gamma Ray Observations}}.
  \href{http://dx.doi.org/10.1103/PhysRevD.88.083521,
  10.1103/PhysRevD.89.049901}{Phys. Rev. {\bf D88} (2013) no.~8, 083521},
  \href{http://arxiv.org/abs/1306.5725}{{\tt arXiv:1306.5725 [astro-ph.HE]}}.
[Erratum: Phys. Rev.D89,no.4,049901(2014)].

\bibitem{Abazajian:2014fta}
K.~N. Abazajian, N.~Canac, S.~Horiuchi, and M.~Kaplinghat, {\em {Astrophysical
  and Dark Matter Interpretations of Extended Gamma-Ray Emission from the
  Galactic Center}}. \href{http://dx.doi.org/10.1103/PhysRevD.90.023526}{Phys.
  Rev. {\bf D90} (2014) no.~2, 023526},
\href{http://arxiv.org/abs/1402.4090}{{\tt arXiv:1402.4090 [astro-ph.HE]}}.

\bibitem{Daylan:2014rsa}
T.~Daylan, D.~P. Finkbeiner, D.~Hooper, T.~Linden, S.~K.~N. Portillo, N.~L.
  Rodd, and T.~R. Slatyer, {\em {The characterization of the gamma-ray signal
  from the central Milky Way: A case for annihilating dark matter}}.
  \href{http://dx.doi.org/10.1016/j.dark.2015.12.005}{Phys. Dark Univ. {\bf 12}
  (2016)  1--23},
\href{http://arxiv.org/abs/1402.6703}{{\tt arXiv:1402.6703 [astro-ph.HE]}}.

\bibitem{Calore:2014xka}
F.~Calore, I.~Cholis, and C.~Weniger, {\em {Background model systematics for
  the Fermi GeV excess}}.
  \href{http://dx.doi.org/10.1088/1475-7516/2015/03/038}{JCAP {\bf 1503} (2015)
   038},
\href{http://arxiv.org/abs/1409.0042}{{\tt arXiv:1409.0042 [astro-ph.CO]}}.

\bibitem{TheFermi-LAT:2015kwa}
{\bf Fermi-LAT} Collaboration, M.~Ajello {\em et al.}, {\em {Fermi-LAT
  Observations of High-Energy $\gamma$-Ray Emission Toward the Galactic
  Center}}. \href{http://dx.doi.org/10.3847/0004-637X/819/1/44}{Astrophys. J.
  {\bf 819} (2016) no.~1, 44},
\href{http://arxiv.org/abs/1511.02938}{{\tt arXiv:1511.02938 [astro-ph.HE]}}.

\bibitem{Petrovic:2014uda}
J.~Petrovic, P.~D. Serpico, and G.~Zaharijaš, {\em {Galactic Center gamma-ray
  "excess" from an active past of the Galactic Centre?}}
  \href{http://dx.doi.org/10.1088/1475-7516/2014/10/052}{JCAP {\bf 1410} (2014)
  no.~10, 052},
\href{http://arxiv.org/abs/1405.7928}{{\tt arXiv:1405.7928 [astro-ph.HE]}}.

\bibitem{Petrovic:2014xra}
J.~Petrovic, P.~D. Serpico, and G.~Zaharijas, {\em {Millisecond pulsars and the
  Galactic Center gamma-ray excess: the importance of luminosity function and
  secondary emission}}.
  \href{http://dx.doi.org/10.1088/1475-7516/2015/02/023}{JCAP {\bf 1502} (2015)
  no.~02, 023},
\href{http://arxiv.org/abs/1411.2980}{{\tt arXiv:1411.2980 [astro-ph.HE]}}.

\bibitem{Mirabal:2012em}
N.~Mirabal, V.~Frias-Martinez, T.~Hassan, and E.~Frias-Martinez, {\em {Fermi's
  Sibyl: Mining the gamma-ray sky for dark matter subhaloes}}.
  \href{http://dx.doi.org/10.1111/j.1745-3933.2012.01287.x}{Mon. Not. Roy.
  Astron. Soc. {\bf 424} (2012)  L64},
\href{http://arxiv.org/abs/1205.4825}{{\tt arXiv:1205.4825 [astro-ph.HE]}}.

\bibitem{Yuan:2014rca}
Q.~Yuan and B.~Zhang, {\em {Millisecond pulsar interpretation of the Galactic
  center gamma-ray excess}}.
  \href{http://dx.doi.org/10.1016/j.jheap.2014.06.001}{JHEAp {\bf 3-4} (2014)
  1--8},
\href{http://arxiv.org/abs/1404.2318}{{\tt arXiv:1404.2318 [astro-ph.HE]}}.

\bibitem{Bartels:2015aea}
R.~Bartels, S.~Krishnamurthy, and C.~Weniger, {\em {Strong support for the
  millisecond pulsar origin of the Galactic center GeV excess}}.
  \href{http://dx.doi.org/10.1103/PhysRevLett.116.051102}{Phys. Rev. Lett. {\bf
  116} (2016) no.~5, 051102},
\href{http://arxiv.org/abs/1506.05104}{{\tt arXiv:1506.05104 [astro-ph.HE]}}.

\bibitem{Lee:2015fea}
S.~K. Lee, M.~Lisanti, B.~R. Safdi, T.~R. Slatyer, and W.~Xue, {\em {Evidence
  for Unresolved $\gamma$-Ray Point Sources in the Inner Galaxy}}.
  \href{http://dx.doi.org/10.1103/PhysRevLett.116.051103}{Phys. Rev. Lett. {\bf
  116} (2016) no.~5, 051103},
\href{http://arxiv.org/abs/1506.05124}{{\tt arXiv:1506.05124 [astro-ph.HE]}}.

\bibitem{Cholis:2015dea}
I.~Cholis, C.~Evoli, F.~Calore, T.~Linden, C.~Weniger, and D.~Hooper, {\em {The
  Galactic Center GeV Excess from a Series of Leptonic Cosmic-Ray Outbursts}}.
  \href{http://dx.doi.org/10.1088/1475-7516/2015/12/005}{JCAP {\bf 1512} (2015)
  no.~12, 005},
\href{http://arxiv.org/abs/1506.05119}{{\tt arXiv:1506.05119 [astro-ph.HE]}}.

\bibitem{Alves:2014yha}
A.~Alves, S.~Profumo, F.~S. Queiroz, and W.~Shepherd, {\em {Effective field
  theory approach to the Galactic Center gamma-ray excess}}.
  \href{http://dx.doi.org/10.1103/PhysRevD.90.115003}{Phys. Rev. {\bf D90}
  (2014) no.~11, 115003},
\href{http://arxiv.org/abs/1403.5027}{{\tt arXiv:1403.5027 [hep-ph]}}.

\bibitem{Calore:2014nla}
F.~Calore, I.~Cholis, C.~McCabe, and C.~Weniger, {\em {A Tale of Tails: Dark
  Matter Interpretations of the Fermi GeV Excess in Light of Background Model
  Systematics}}. \href{http://dx.doi.org/10.1103/PhysRevD.91.063003}{Phys. Rev.
  {\bf D91} (2015) no.~6, 063003},
\href{http://arxiv.org/abs/1411.4647}{{\tt arXiv:1411.4647 [hep-ph]}}.

\bibitem{Agrawal:2014oha}
P.~Agrawal, B.~Batell, P.~J. Fox, and R.~Harnik, {\em {WIMPs at the Galactic
  Center}}. \href{http://dx.doi.org/10.1088/1475-7516/2015/05/011}{JCAP {\bf
  1505} (2015)  011},
\href{http://arxiv.org/abs/1411.2592}{{\tt arXiv:1411.2592 [hep-ph]}}.

\bibitem{Okada:2013bna}
N.~Okada and O.~Seto, {\em {Gamma ray emission in Fermi bubbles and Higgs
  portal dark matter}}.
  \href{http://dx.doi.org/10.1103/PhysRevD.89.043525}{Phys. Rev. {\bf D89}
  (2014) no.~4, 043525},
\href{http://arxiv.org/abs/1310.5991}{{\tt arXiv:1310.5991 [hep-ph]}}.

\bibitem{Basak:2014sza}
T.~Mondal and T.~Basak, {\em {Class of Higgs-portal Dark Matter models in the
  light of gamma-ray excess from Galactic center}}.
  \href{http://dx.doi.org/10.1016/j.physletb.2015.03.055}{Phys. Lett. {\bf
  B744} (2015)  208--212},
\href{http://arxiv.org/abs/1405.4877}{{\tt arXiv:1405.4877 [hep-ph]}}.

\bibitem{Cline:2013gha}
J.~M. Cline, K.~Kainulainen, P.~Scott, and C.~Weniger, {\em {Update on scalar
  singlet dark matter}}. \href{http://dx.doi.org/10.1103/PhysRevD.92.039906,
  10.1103/PhysRevD.88.055025}{Phys. Rev. {\bf D88} (2013)  055025},
  \href{http://arxiv.org/abs/1306.4710}{{\tt arXiv:1306.4710 [hep-ph]}}.
[Erratum: Phys. Rev.D92,no.3,039906(2015)].

\bibitem{Duerr:2015bea}
M.~Duerr, P.~Fileviez~PŽrez, and J.~Smirnov, {\em {Gamma-Ray Excess and the
  Minimal Dark Matter Model}}.
  \href{http://dx.doi.org/10.1007/JHEP06(2016)008}{JHEP {\bf 06} (2016)  008},
\href{http://arxiv.org/abs/1510.07562}{{\tt arXiv:1510.07562 [hep-ph]}}.

\bibitem{Gonderinger:2009jp}
M.~Gonderinger, Y.~Li, H.~Patel, and M.~J. Ramsey-Musolf, {\em {Vacuum
  Stability, Perturbativity, and Scalar Singlet Dark Matter}}.
  \href{http://dx.doi.org/10.1007/JHEP01(2010)053}{JHEP {\bf 01} (2010)  053},
\href{http://arxiv.org/abs/0910.3167}{{\tt arXiv:0910.3167 [hep-ph]}}.

\bibitem{Bernal:2015xba}
N.~Bernal and X.~Chu, {\em {$Z_2$ SIMP Dark Matter}}.
  \href{http://dx.doi.org/10.1088/1475-7516/2016/01/006}{JCAP {\bf 1601} (2016)
   006},
\href{http://arxiv.org/abs/1510.08527}{{\tt arXiv:1510.08527 [hep-ph]}}.

\bibitem{Beniwal:2015sdl}
A.~Beniwal, F.~Rajec, C.~Savage, P.~Scott, C.~Weniger, M.~White, and A.~G.
  Williams, {\em {Combined analysis of effective Higgs portal dark matter
  models}}. \href{http://dx.doi.org/10.1103/PhysRevD.93.115016}{Phys. Rev. {\bf
  D93} (2016) no.~11, 115016},
\href{http://arxiv.org/abs/1512.06458}{{\tt arXiv:1512.06458 [hep-ph]}}.

\bibitem{Djouadi:2011aa}
A.~Djouadi, O.~Lebedev, Y.~Mambrini, and J.~Quevillon, {\em {Implications of
  LHC searches for Higgs--portal dark matter}}.
  \href{http://dx.doi.org/10.1016/j.physletb.2012.01.062}{Phys. Lett. {\bf
  B709} (2012)  65--69},
\href{http://arxiv.org/abs/1112.3299}{{\tt arXiv:1112.3299 [hep-ph]}}.

\bibitem{Cheung:2012xb}
K.~Cheung, Y.-L.~S. Tsai, P.-Y. Tseng, T.-C. Yuan, and A.~Zee, {\em {Global
  Study of the Simplest Scalar Phantom Dark Matter Model}}.
  \href{http://dx.doi.org/10.1088/1475-7516/2012/10/042}{JCAP {\bf 1210} (2012)
   042},
\href{http://arxiv.org/abs/1207.4930}{{\tt arXiv:1207.4930 [hep-ph]}}.

\bibitem{Djouadi:2012zc}
A.~Djouadi, A.~Falkowski, Y.~Mambrini, and J.~Quevillon, {\em {Direct Detection
  of Higgs-Portal Dark Matter at the LHC}}.
  \href{http://dx.doi.org/10.1140/epjc/s10052-013-2455-1}{Eur. Phys. J. {\bf
  C73} (2013) no.~6, 2455},
\href{http://arxiv.org/abs/1205.3169}{{\tt arXiv:1205.3169 [hep-ph]}}.

\bibitem{Endo:2014cca}
M.~Endo and Y.~Takaesu, {\em {Heavy WIMP through Higgs portal at the LHC}}.
  \href{http://dx.doi.org/10.1016/j.physletb.2015.02.042}{Phys. Lett. {\bf
  B743} (2015)  228--234},
\href{http://arxiv.org/abs/1407.6882}{{\tt arXiv:1407.6882 [hep-ph]}}.

\bibitem{Craig:2014lda}
N.~Craig, H.~K. Lou, M.~McCullough, and A.~Thalapillil, {\em {The Higgs Portal
  Above Threshold}}. \href{http://dx.doi.org/10.1007/JHEP02(2016)127}{JHEP {\bf
  02} (2016)  127},
\href{http://arxiv.org/abs/1412.0258}{{\tt arXiv:1412.0258 [hep-ph]}}.

\bibitem{Han:2016gyy}
H.~Han, J.~M. Yang, Y.~Zhang, and S.~Zheng, {\em {Collider Signatures of
  Higgs-portal Scalar Dark Matter}}.
  \href{http://dx.doi.org/10.1016/j.physletb.2016.03.010}{Phys. Lett. {\bf
  B756} (2016)  109--112},
\href{http://arxiv.org/abs/1601.06232}{{\tt arXiv:1601.06232 [hep-ph]}}.

\bibitem{Mambrini:2012ue}
Y.~Mambrini, M.~H.~G. Tytgat, G.~Zaharijas, and B.~Zaldivar, {\em
  {Complementarity of Galactic radio and collider data in constraining WIMP
  dark matter models}}.
  \href{http://dx.doi.org/10.1088/1475-7516/2012/11/038}{JCAP {\bf 1211} (2012)
   038},
\href{http://arxiv.org/abs/1206.2352}{{\tt arXiv:1206.2352 [hep-ph]}}.

\bibitem{Feng:2014vea}
L.~Feng, S.~Profumo, and L.~Ubaldi, {\em {Closing in on singlet scalar dark
  matter: LUX, invisible Higgs decays and gamma-ray lines}}.
  \href{http://dx.doi.org/10.1007/JHEP03(2015)045}{JHEP {\bf 03} (2015)  045},
\href{http://arxiv.org/abs/1412.1105}{{\tt arXiv:1412.1105 [hep-ph]}}.

\bibitem{Duerr:2015mva}
M.~Duerr, P.~Fileviez~Perez, and J.~Smirnov, {\em {Scalar Singlet Dark Matter
  and Gamma Lines}}.
  \href{http://dx.doi.org/10.1016/j.physletb.2015.10.034}{Phys. Lett. {\bf
  B751} (2015)  119--122},
\href{http://arxiv.org/abs/1508.04418}{{\tt arXiv:1508.04418 [hep-ph]}}.

\bibitem{Duerr:2015aka}
M.~Duerr, P.~Fileviez~Perez, and J.~Smirnov, {\em {Scalar Dark Matter: Direct
  vs. Indirect Detection}}.
\href{http://arxiv.org/abs/1509.04282}{{\tt arXiv:1509.04282 [hep-ph]}}.

\bibitem{Han:2015hda}
H.~Han and S.~Zheng, {\em {New Constraints on Higgs-portal Scalar Dark
  Matter}}. \href{http://dx.doi.org/10.1007/JHEP12(2015)044}{JHEP {\bf 12}
  (2015)  044},
\href{http://arxiv.org/abs/1509.01765}{{\tt arXiv:1509.01765 [hep-ph]}}.

\bibitem{Kahlhoefer:2015jma}
F.~Kahlhoefer and J.~McDonald, {\em {WIMP Dark Matter and Unitarity-Conserving
  Inflation via a Gauge Singlet Scalar}}.
  \href{http://dx.doi.org/10.1088/1475-7516/2015/11/015}{JCAP {\bf 1511} (2015)
  no.~11, 015},
\href{http://arxiv.org/abs/1507.03600}{{\tt arXiv:1507.03600 [astro-ph.CO]}}.

\bibitem{Alvares:2012qv}
J.~D. Ruiz-Alvarez, C.~A. de~S.~Pires, F.~S. Queiroz, D.~Restrepo, and P.~S.
  Rodrigues~da Silva, {\em {On the Connection of Gamma-Rays, Dark Matter and
  Higgs Searches at LHC}}.
  \href{http://dx.doi.org/10.1103/PhysRevD.86.075011}{Phys. Rev. {\bf D86}
  (2012)  075011},
\href{http://arxiv.org/abs/1206.5779}{{\tt arXiv:1206.5779 [hep-ph]}}.

\bibitem{Ghorbani:2014gka}
K.~Ghorbani and H.~Ghorbani, {\em {Scalar split WIMPs in future direct
  detection experiments}}.
  \href{http://dx.doi.org/10.1103/PhysRevD.93.055012}{Phys. Rev. {\bf D93}
  (2016) no.~5, 055012},
\href{http://arxiv.org/abs/1501.00206}{{\tt arXiv:1501.00206 [hep-ph]}}.

\bibitem{Wang:2014elb}
L.~Wang and X.-F. Han, {\em {A simplified 2HDM with a scalar dark matter and
  the galactic center gamma-ray excess}}.
  \href{http://dx.doi.org/10.1016/j.physletb.2014.11.016}{Phys. Lett. {\bf
  B739} (2014)  416--420},
\href{http://arxiv.org/abs/1406.3598}{{\tt arXiv:1406.3598 [hep-ph]}}.

\bibitem{Kim:2016csm}
Y.~G. Kim, K.~Y. Lee, C.~B. Park, and S.~Shin, {\em {Secluded singlet fermionic
  dark matter driven by the Fermi gamma-ray excess}}.
  \href{http://dx.doi.org/10.1103/PhysRevD.93.075023}{Phys. Rev. {\bf D93}
  (2016) no.~7, 075023},
\href{http://arxiv.org/abs/1601.05089}{{\tt arXiv:1601.05089 [hep-ph]}}.

\bibitem{Alloul:2013bka}
A.~Alloul, N.~D. Christensen, C.~Degrande, C.~Duhr, and B.~Fuks, {\em
  {FeynRules 2.0 - A complete toolbox for tree-level phenomenology}}.
  \href{http://dx.doi.org/10.1016/j.cpc.2014.04.012}{Comput.Phys.Commun. {\bf
  185} (2014)  2250--2300},
\href{http://arxiv.org/abs/1310.1921}{{\tt arXiv:1310.1921 [hep-ph]}}.

\bibitem{Belanger:2014vza}
G.~BŽlanger, F.~Boudjema, A.~Pukhov, and A.~Semenov, {\em {micrOMEGAs4.1: two
  dark matter candidates}}.
  \href{http://dx.doi.org/10.1016/j.cpc.2015.03.003}{Comput. Phys. Commun. {\bf
  192} (2015)  322--329},
\href{http://arxiv.org/abs/1407.6129}{{\tt arXiv:1407.6129 [hep-ph]}}.

\bibitem{Belyaev:2012qa}
A.~Belyaev, N.~D. Christensen, and A.~Pukhov, {\em {CalcHEP 3.4 for collider
  physics within and beyond the Standard Model}}.
  \href{http://dx.doi.org/10.1016/j.cpc.2013.01.014}{Comput.Phys.Commun. {\bf
  184} (2013)  1729--1769},
\href{http://arxiv.org/abs/1207.6082}{{\tt arXiv:1207.6082 [hep-ph]}}.

\bibitem{Shifman:1979eb}
M.~A. Shifman, A.~I. Vainshtein, M.~B. Voloshin, and V.~I. Zakharov, {\em
  {Low-Energy Theorems for Higgs Boson Couplings to Photons}}. Sov. J. Nucl.
  Phys. {\bf 30} (1979)  711--716.
[Yad. Fiz.30,1368(1979)].

\bibitem{Dawson:1990zj}
S.~Dawson, {\em {Radiative corrections to Higgs boson production}}.
\href{http://dx.doi.org/10.1016/0550-3213(91)90061-2}{Nucl. Phys. {\bf B359}
  (1991)  283--300}.

\bibitem{Spira:1995rr}
M.~Spira, A.~Djouadi, D.~Graudenz, and P.~M. Zerwas, {\em {Higgs boson
  production at the LHC}}.
  \href{http://dx.doi.org/10.1016/0550-3213(95)00379-7}{Nucl. Phys. {\bf B453}
  (1995)  17--82},
\href{http://arxiv.org/abs/hep-ph/9504378}{{\tt arXiv:hep-ph/9504378
  [hep-ph]}}.

\bibitem{Dittmaier:2011ti}
{\bf LHC Higgs Cross Section Working Group} Collaboration, S.~Dittmaier {\em et
  al.}, {\em {Handbook of LHC Higgs Cross Sections: 1. Inclusive Observables}}.
\href{http://arxiv.org/abs/1101.0593}{{\tt arXiv:1101.0593 [hep-ph]}}.

\bibitem{Aad:2015zhl}
{\bf ATLAS, CMS} Collaboration, G.~Aad {\em et al.}, {\em {Combined Measurement
  of the Higgs Boson Mass in $pp$ Collisions at $\sqrt{s}=7$ and 8 TeV with the
  ATLAS and CMS Experiments}}.
  \href{http://dx.doi.org/10.1103/PhysRevLett.114.191803}{Phys. Rev. Lett. {\bf
  114} (2015)  191803},
\href{http://arxiv.org/abs/1503.07589}{{\tt arXiv:1503.07589 [hep-ex]}}.

\bibitem{Lacroix:2014eea}
T.~Lacroix, C.~Boehm, and J.~Silk, {\em {Fitting the Fermi-LAT GeV excess: On
  the importance of including the propagation of electrons from dark matter}}.
  \href{http://dx.doi.org/10.1103/PhysRevD.90.043508}{Phys. Rev. {\bf D90}
  (2014) no.~4, 043508},
\href{http://arxiv.org/abs/1403.1987}{{\tt arXiv:1403.1987 [astro-ph.HE]}}.

\bibitem{SjMrSk07}
T.~Sjostrand, S.~Mrenna, and P.~Z. Skands, {\em {A Brief Introduction to PYTHIA
  8.1}}. \href{http://dx.doi.org/10.1016/j.cpc.2008.01.036}{Comput.Phys.Commun.
  {\bf 178} (2008)  852--867},
\href{http://arxiv.org/abs/0710.3820}{{\tt arXiv:0710.3820 [hep-ph]}}.

\bibitem{Alwall:2011uj}
J.~Alwall, M.~Herquet, F.~Maltoni, O.~Mattelaer, and T.~Stelzer, {\em {MadGraph
  5 : Going Beyond}}. \href{http://dx.doi.org/10.1007/JHEP06(2011)128}{JHEP
  {\bf 1106} (2011)  128},
\href{http://arxiv.org/abs/1106.0522}{{\tt arXiv:1106.0522 [hep-ph]}}.

\bibitem{Alwall:2014hca}
J.~Alwall, R.~Frederix, S.~Frixione, V.~Hirschi, F.~Maltoni, {\em et al.}, {\em
  {The automated computation of tree-level and next-to-leading order
  differential cross sections, and their matching to parton shower
  simulations}}. \href{http://dx.doi.org/10.1007/JHEP07(2014)079}{JHEP {\bf
  1407} (2014)  079},
\href{http://arxiv.org/abs/1405.0301}{{\tt arXiv:1405.0301 [hep-ph]}}.

\bibitem{Sjostrand:2006za}
T.~Sjostrand, S.~Mrenna, and P.~Z. Skands, {\em {PYTHIA 6.4 Physics and
  Manual}}. \href{http://dx.doi.org/10.1088/1126-6708/2006/05/026}{JHEP {\bf
  0605} (2006)  026},
\href{http://arxiv.org/abs/hep-ph/0603175}{{\tt arXiv:hep-ph/0603175
  [hep-ph]}}.

\bibitem{Cirelli:2010xx}
M.~Cirelli, G.~Corcella, A.~Hektor, G.~Hutsi, M.~Kadastik, P.~Panci, M.~Raidal,
  F.~Sala, and A.~Strumia, {\em {PPPC 4 DM ID: A Poor Particle Physicist
  Cookbook for Dark Matter Indirect Detection}}.
  \href{http://dx.doi.org/10.1088/1475-7516/2012/10/E01,
  10.1088/1475-7516/2011/03/051}{JCAP {\bf 1103} (2011)  051},
  \href{http://arxiv.org/abs/1012.4515}{{\tt arXiv:1012.4515 [hep-ph]}}.
[Erratum: JCAP1210,E01(2012)].

\bibitem{Caron:2015wda}
A.~Achterberg, S.~Amoroso, S.~Caron, L.~Hendriks, R.~Ruiz~de Austri, and
  C.~Weniger, {\em {A description of the Galactic Center excess in the Minimal
  Supersymmetric Standard Model}}.
  \href{http://dx.doi.org/10.1088/1475-7516/2015/08/006}{JCAP {\bf 1508} (2015)
  no.~08, 006},
\href{http://arxiv.org/abs/1502.05703}{{\tt arXiv:1502.05703 [hep-ph]}}.

\bibitem{Navarro:1995iw}
J.~F. Navarro, C.~S. Frenk, and S.~D. White, {\em {The Structure of cold dark
  matter halos}}. \href{http://dx.doi.org/10.1086/177173}{Astrophys.J. {\bf
  462} (1996)  563--575},
\href{http://arxiv.org/abs/astro-ph/9508025}{{\tt arXiv:astro-ph/9508025
  [astro-ph]}}.

\bibitem{Nesti:2013uwa}
F.~Nesti and P.~Salucci, {\em {The Dark Matter halo of the Milky Way, AD
  2013}}. \href{http://dx.doi.org/10.1088/1475-7516/2013/07/016}{JCAP {\bf
  1307} (2013)  016},
\href{http://arxiv.org/abs/1304.5127}{{\tt arXiv:1304.5127 [astro-ph.GA]}}.

\bibitem{Catena:2009mf}
R.~Catena and P.~Ullio, {\em {A novel determination of the local dark matter
  density}}. \href{http://dx.doi.org/10.1088/1475-7516/2010/08/004}{JCAP {\bf
  1008} (2010)  004},
\href{http://arxiv.org/abs/0907.0018}{{\tt arXiv:0907.0018 [astro-ph.CO]}}.

\bibitem{Pato:2015dua}
M.~Pato, F.~Iocco, and G.~Bertone, {\em {Dynamical constraints on the dark
  matter distribution in the Milky Way}}.
  \href{http://dx.doi.org/10.1088/1475-7516/2015/12/001}{JCAP {\bf 1512} (2015)
  no.~12, 001},
\href{http://arxiv.org/abs/1504.06324}{{\tt arXiv:1504.06324 [astro-ph.GA]}}.

\bibitem{McMillan:2011wd}
P.~J. McMillan, {\em {Mass models of the Milky Way}}.
  \href{http://dx.doi.org/10.1111/j.1365-2966.2011.18564.x}{Mon.Not.Roy.Astron.Soc.
  {\bf 414} (2011)  2446--2457},
\href{http://arxiv.org/abs/1102.4340}{{\tt arXiv:1102.4340 [astro-ph.GA]}}.

\bibitem{Salucci:2010qr}
P.~Salucci, F.~Nesti, G.~Gentile, and C.~Martins, {\em {The dark matter density
  at the Sun's location}}.
  \href{http://dx.doi.org/10.1051/0004-6361/201014385}{Astron.Astrophys. {\bf
  523} (2010)  A83},
\href{http://arxiv.org/abs/1003.3101}{{\tt arXiv:1003.3101 [astro-ph.GA]}}.

\bibitem{Read:2014qva}
J.~I. Read, {\em {The Local Dark Matter Density}}.
  \href{http://dx.doi.org/10.1088/0954-3899/41/6/063101}{J. Phys. {\bf G41}
  (2014)  063101},
\href{http://arxiv.org/abs/1404.1938}{{\tt arXiv:1404.1938 [astro-ph.GA]}}.

\bibitem{Feroz:2008xx}
F.~Feroz, M.~P. Hobson, and M.~Bridges, {\em {MultiNest: an efficient and
  robust Bayesian inference tool for cosmology and particle physics}}.
  \href{http://dx.doi.org/10.1111/j.1365-2966.2009.14548.x}{Mon. Not. Roy.
  Astron. Soc. {\bf 398} (2009)  1601--1614},
\href{http://arxiv.org/abs/0809.3437}{{\tt arXiv:0809.3437 [astro-ph]}}.

\bibitem{Feroz:2013hea}
F.~Feroz, M.~P. Hobson, E.~Cameron, and A.~N. Pettitt, {\em {Importance Nested
  Sampling and the MultiNest Algorithm}}.
\href{http://arxiv.org/abs/1306.2144}{{\tt arXiv:1306.2144 [astro-ph.IM]}}.

\bibitem{Rolke:2004mj}
W.~A. Rolke, A.~M. Lopez, and J.~Conrad, {\em {Limits and confidence intervals
  in the presence of nuisance parameters}}.
  \href{http://dx.doi.org/10.1016/j.nima.2005.05.068}{Nucl. Instrum. Meth. {\bf
  A551} (2005)  493--503},
\href{http://arxiv.org/abs/physics/0403059}{{\tt arXiv:physics/0403059
  [physics]}}.

\bibitem{Slatyer:2015jla}
T.~R. Slatyer, {\em {Indirect dark matter signatures in the cosmic dark ages.
  I. Generalizing the bound on s-wave dark matter annihilation from Planck
  results}}. \href{http://dx.doi.org/10.1103/PhysRevD.93.023527}{Phys. Rev.
  {\bf D93} (2016) no.~2, 023527},
\href{http://arxiv.org/abs/1506.03811}{{\tt arXiv:1506.03811 [hep-ph]}}.

\bibitem{Aad:2015pla}
{\bf ATLAS} Collaboration, G.~Aad {\em et al.}, {\em {Constraints on new
  phenomena via Higgs boson couplings and invisible decays with the ATLAS
  detector}}. \href{http://dx.doi.org/10.1007/JHEP11(2015)206}{JHEP {\bf 11}
  (2015)  206},
\href{http://arxiv.org/abs/1509.00672}{{\tt arXiv:1509.00672 [hep-ex]}}.

\bibitem{Kanemura:2010sh}
S.~Kanemura, S.~Matsumoto, T.~Nabeshima, and N.~Okada, {\em {Can WIMP Dark
  Matter overcome the Nightmare Scenario?}}
  \href{http://dx.doi.org/10.1103/PhysRevD.82.055026}{Phys. Rev. {\bf D82}
  (2010)  055026},
\href{http://arxiv.org/abs/1005.5651}{{\tt arXiv:1005.5651 [hep-ph]}}.

\bibitem{Akerib:2013tjd}
{\bf LUX} Collaboration, D.~S. Akerib {\em et al.}, {\em {First results from
  the LUX dark matter experiment at the Sanford Underground Research
  Facility}}. \href{http://dx.doi.org/10.1103/PhysRevLett.112.091303}{Phys.
  Rev. Lett. {\bf 112} (2014)  091303},
\href{http://arxiv.org/abs/1310.8214}{{\tt arXiv:1310.8214 [astro-ph.CO]}}.

\bibitem{Savage:2015xta}
C.~Savage, A.~Scaffidi, M.~White, and A.~G. Williams, {\em {LUX likelihood and
  limits on spin-independent and spin-dependent WIMP couplings with LUXCalc}}.
  \href{http://dx.doi.org/10.1103/PhysRevD.92.103519}{Phys. Rev. {\bf D92}
  (2015) no.~10, 103519},
\href{http://arxiv.org/abs/1502.02667}{{\tt arXiv:1502.02667 [hep-ph]}}.

\bibitem{Aprile:2015uzo}
{\bf XENON} Collaboration, E.~Aprile {\em et al.}, {\em {Physics reach of the
  XENON1T dark matter experiment}}.
  \href{http://dx.doi.org/10.1088/1475-7516/2016/04/027}{JCAP {\bf 1604} (2016)
  no.~04, 027},
\href{http://arxiv.org/abs/1512.07501}{{\tt arXiv:1512.07501
  [physics.ins-det]}}.

\bibitem{Baudis:2012bc}
{\bf DARWIN Consortium} Collaboration, L.~Baudis, {\em {DARWIN: dark matter
  WIMP search with noble liquids}}.
  \href{http://dx.doi.org/10.1088/1742-6596/375/1/012028}{J. Phys. Conf. Ser.
  {\bf 375} (2012)  012028},
\href{http://arxiv.org/abs/1201.2402}{{\tt arXiv:1201.2402 [astro-ph.IM]}}.

\bibitem{Ackermann:2015zua}
{\bf Fermi-LAT} Collaboration, M.~Ackermann {\em et al.}, {\em {Searching for
  Dark Matter Annihilation from Milky Way Dwarf Spheroidal Galaxies with Six
  Years of Fermi Large Area Telescope Data}}.
  \href{http://dx.doi.org/10.1103/PhysRevLett.115.231301}{Phys. Rev. Lett. {\bf
  115} (2015) no.~23, 231301},
\href{http://arxiv.org/abs/1503.02641}{{\tt arXiv:1503.02641 [astro-ph.HE]}}.

\bibitem{Ahnen:2016qkx}
{\bf Fermi-LAT, MAGIC} Collaboration, M.~L. Ahnen {\em et al.}, {\em {Limits to
  dark matter annihilation cross-section from a combined analysis of MAGIC and
  Fermi-LAT observations of dwarf satellite galaxies}}.
  \href{http://dx.doi.org/10.1088/1475-7516/2016/02/039}{JCAP {\bf 1602} (2016)
  no.~02, 039},
\href{http://arxiv.org/abs/1601.06590}{{\tt arXiv:1601.06590 [astro-ph.HE]}}.

\bibitem{Ackermann:2015lka}
{\bf Fermi-LAT} Collaboration, M.~Ackermann {\em et al.}, {\em {Updated search
  for spectral lines from Galactic dark matter interactions with pass 8 data
  from the Fermi Large Area Telescope}}.
  \href{http://dx.doi.org/10.1103/PhysRevD.91.122002}{Phys. Rev. {\bf D91}
  (2015) no.~12, 122002},
\href{http://arxiv.org/abs/1506.00013}{{\tt arXiv:1506.00013 [astro-ph.HE]}}.

\bibitem{Ade:2015xua}
{\bf Planck} Collaboration, P.~A.~R. Ade {\em et al.}, {\em {Planck 2015
  results. XIII. Cosmological parameters}}.
\href{http://arxiv.org/abs/1502.01589}{{\tt arXiv:1502.01589 [astro-ph.CO]}}.

\bibitem{Bonnivard:2015xpq}
V.~Bonnivard {\em et al.}, {\em {Dark matter annihilation and decay in dwarf
  spheroidal galaxies: The classical and ultrafaint dSphs}}.
  \href{http://dx.doi.org/10.1093/mnras/stv1601}{Mon. Not. Roy. Astron. Soc.
  {\bf 453} (2015) no.~1, 849--867},
\href{http://arxiv.org/abs/1504.02048}{{\tt arXiv:1504.02048 [astro-ph.HE]}}.

\bibitem{Consortium:2010bc}
{\bf CTA Consortium} Collaboration, M.~Actis {\em et al.}, {\em {Design
  concepts for the Cherenkov Telescope Array CTA: An advanced facility for
  ground-based high-energy gamma-ray astronomy}}.
  \href{http://dx.doi.org/10.1007/s10686-011-9247-0}{Exper. Astron. {\bf 32}
  (2011)  193--316},
\href{http://arxiv.org/abs/1008.3703}{{\tt arXiv:1008.3703 [astro-ph.IM]}}.

\bibitem{Silverwood:2014yza}
H.~Silverwood, C.~Weniger, P.~Scott, and G.~Bertone, {\em {A realistic
  assessment of the CTA sensitivity to dark matter annihilation}}.
  \href{http://dx.doi.org/10.1088/1475-7516/2015/03/055}{JCAP {\bf 1503} (2015)
  no.~03, 055},
\href{http://arxiv.org/abs/1408.4131}{{\tt arXiv:1408.4131 [astro-ph.HE]}}.

\bibitem{Abbott:2005bi}
{\bf Dark Energy Survey} Collaboration, T.~Abbott {\em et al.}, {\em {The dark
  energy survey}}.
\href{http://arxiv.org/abs/astro-ph/0510346}{{\tt arXiv:astro-ph/0510346
  [astro-ph]}}.

\bibitem{Ivezic:2008fe}
{\bf LSST} Collaboration, Z.~Ivezic, J.~A. Tyson, R.~Allsman, J.~Andrew, and
  R.~Angel, {\em {LSST: from Science Drivers to Reference Design and
  Anticipated Data Products}}.
\href{http://arxiv.org/abs/0805.2366}{{\tt arXiv:0805.2366 [astro-ph]}}.

\end{thebibliography}\endgroup

\end{document}